\documentclass{optica-article}
\usepackage{tabularx}
\usepackage{array}
\newcolumntype{Y}{>{\raggedright\arraybackslash}X}
\journal{opticajournal}

\articletype{Research Article}

\usepackage{lineno}

\usepackage{amsmath,amssymb}
\usepackage{siunitx}
\usepackage{multirow}
\usepackage{booktabs}
\usepackage{graphicx}
\usepackage{subcaption}

\begin{document}

\title{Design Guidelines for Plasmon-Enhanced CsSn$_x$Ge$_{1-x}$I$_3$ Perovskite LEDs: A DFT-Informed FDTD Study}

\author{Shoumik Debnath,\authormark{1} Sudipta Saha,\authormark{1} Khondokar Zahin,\authormark{1} Ying Yin Tsui,\authormark{2} and Md. Zahurul Islam\authormark{1,*}}

\address{\authormark{1}Department of Electrical and Electronic Engineering, Bangladesh University of Engineering and Technology, Dhaka 1000, Bangladesh\\
\authormark{2}Department of Electrical and Computer Engineering, University of Alberta, Edmonton, AB T6G 2H5, Canada}

\email{\authormark{*}mdzahurulislam@eee.buet.ac.bd}

\begin{abstract*}
CsSn$_x$Ge$_{1-x}$I$_3$ as lead-free perovskites are promising for next generation NIR emitting perovskite LEDs due to their tunable bandgaps and stability. However, they suffer from poor light extraction efficiency, and accurate composition-specific optical data for these materials remain scarce. This study presents a DFT-FDTD framework to optimize light extraction via compositional tuning and plasmonic enhancement. First, DFT calculations were performed to obtain composition-specific complex refractive index and extinction coefficient values for $x = 0, 0.25, 0.5, 0.75$, and $1$. Results show bandgap increased from 1.331 eV for CsSnI$_3$ to 1.927 eV for CsGeI$_3$ with increasing Ge content, while refractive index ranges from 2.2 to 2.6 across compositions. These optical constants were then used as inputs for FDTD simulations of a PeLED structure with optimized Au/SiO$_2$ core-shell nanorods for plasmonic enhancement. A 12.1-fold Purcell enhancement was achieved for CsSn$_{0.25}$Ge$_{0.75}$I$_3$, while light extraction efficiency reached 25\% for CsSn$_{0.5}$Ge$_{0.5}$I$_3$. LEE enhancement of 36\% was obtained for CsSnI$_3$, and spectral overlap between emitter and plasmon resonance reached 96\% for Sn-rich compositions. Design guidelines indicate CsSn$_{0.5}$Ge$_{0.5}$I$_3$ offers optimal balance of extraction efficiency (25\%), Purcell enhancement (5.3$\times$), spectral overlap (93\%), and oxidation  stability  for wearable and flexible optoelectronic applications, while CsSn$_{0.25}$Ge$_{0.75}$I$_3$ is recommended for applications prioritizing spontaneous emission rate.
\end{abstract*}

\section{Introduction}

PeLEDs have rapidly emerged as a promising candidate for next-generation light sources, including applications in displays~\cite{Cho2024,Liu2024a,Rogalski2025}, wearable electronics, and flexible optoelectronics~\cite{Chen2024,Xiong2025}. This rapid advancement has been made possible because of the exceptional optoelectronic properties of metal halide perovskites (MHPs), including high absorption coefficients, solution processability, and tunable emission characteristics~\cite{Lu2024}. The external quantum efficiency (EQE) of these devices has surged from less than 1\% in early demonstrations~\cite{Sanchez-Diaz2024} to over 28\% for red~\cite{Kong2024}, 30\% for green~\cite{Li2024a}, and 23\% for blue emitters~\cite{Nong2024}. This positions PeLEDs as serious competitors to already established organic LED (OLED) technology~\cite{Cho2024}.

Apart from these efficiency gains, PeLEDs offer several other distinct advantages that make them attractive for commercial usage. These advantages, combined with narrow emission linewidths (FWHM typically <20~nm), bandgap tunability across visible to NIR wavelengths, and low-cost solution processing~\cite{Ma2025,Xiao2024,Dong2020,Yang2022,Liu2024b,Ko2025}, position PeLEDs as strong candidates for next-generation displays and flexible optoelectronics.

Although EQE has reached up to 32\%~\cite{Li2024b}, the inherent toxicity of Pb is a major barrier to commercialization of these optoelectronic devices~\cite{Li2024a,Lopez-Fernandez2024,Zhang2021,Chen2021}. So, developing lead-free alternatives is a research priority to ensure environmental safety and sustainability~\cite{Chen2024,Yang2022,Kar2021}. Also, for commercial usage, operational lifetimes should exceed 10,000 hours and manufacturing costs should remain below USD 100~m$^{-2}$~\cite{Zhang2025}. These targets are still unachieved for both lead-based and lead-free systems. This leaves significant room for innovation in both materials engineering and device architecture optimization.

The search for Pb substitutes has focused on tin (Sn$^{2+}$) and germanium (Ge$^{2+}$), which share comparable ionic radii and valence states with Pb$^{2+}$ ~\cite{Lopez-Fernandez2024,Kar2021}. Sn-based PeLEDs have demonstrated external quantum efficiencies (EQEs) up to 20.29\% with narrow emission bandwidth of 24.9~nm ~\cite{Xiao2024}, while Ge--Pb mixed PeLEDs have exhibited photoluminescence quantum yields (PLQYs) of nearly 71\% and EQEs around 13\%~\cite{Yang2021}. Sn-based MHPs exhibit excellent optoelectronic properties with direct bandgaps ranging from 1.2~eV to 1.4~eV, emitting in the near-infrared (NIR) region~\cite{Lopez-Fernandez2024}.

However, both Sn- and Ge-based MHPs suffer from oxidation-induced instability, where Sn$^{2+}$ and Ge$^{2+}$ ions readily oxidize to Sn$^{4+}$ and Ge$^{4+}$, forming deep defects that degrade long-term performance~\cite{Xiao2024,Lopez-Fernandez2024}. This intrinsic chemical instability of Sn$^{2+}$ arises from its relatively low Sn$^{2+}$/Sn$^{4+}$ redox potential, causing spontaneous oxidation in both precursor solutions and deposited thin films~\cite{Wang2024}.

Partial substitution of Sn$^{2+}$ with Ge$^{2+}$ has proven particularly effective in mitigating this instability~\cite{Aftab2024, Kore2024}. Ge alloying forms a native GeO$_x$ layer at the perovskite surface, which acts as a protective barrier against environmental factors such as moisture and oxygen~\cite{Chen2018}. CsSn$_{0.5}$Ge$_{0.5}$I$_3$-based devices have demonstrated stable operation with less than 10\% efficiency degradation over 500 hours of continuous operation~\cite{Chen2018}, representing a significant advancement in operational stability for lead-free systems. This concept, first demonstrated in perovskite photovoltaics, has led to simultaneous improvements in both power conversion efficiency and long-term stability~\cite{Lopez-Fernandez2024}. Compositionally tuned CsSn$_x$Ge$_{1-x}$I$_3$ systems therefore represent a promising alternative to lead-based perovskites for next-generation optoelectronic devices.

However, the high refractive index ($\sim$2.5) of these perovskites limits the outcoupling efficiency to approximately 8\%~\cite{Cho2024,Rahimi2024}. This poor outcoupling happens because the perovskite's high refractive index traps most of the light. Studies show that nearly 80\% of the generated photons never escape the device~\cite{Zhao2023}. The primary reason behind it is total internal reflection at material interfaces. This alone accounts for nearly 56\% of total radiative power loss~\cite{Rahimi2024,Liu2024c}. So, enhancing LEE is a critical step towards realizing highly efficient lead-free PeLEDs.

To model light extraction properly, we need accurate wavelength-dependent $n$ and $k$ values for the perovskite emitter~\cite{Shi2018,Zhang2020}. Existing optical simulations often reuse generic or oversimplified $n$, $k$ datasets resulting in substantial inaccuracies. Generic optical constants and purely geometric models cause inaccuracies in capturing the complex wave-optical behavior of light~\cite{Futscher2017}. Such inaccuracies lead to spectral mismatch between the emitter's photoluminescence and the device's cavity modes. As a result, LEE predictions become unreliable and far-field radiation pattern characteristics become distorted~\cite{Shi2018}.

First-principles density functional theory (DFT) calculations are a robust approach to counter this problem. DFT calculations determine optimized geometrical structures~\cite{Xiong2025} along with accurate electronic structures and optical properties~\cite{Li2024c}. This methodology has been widely used to investigate optical and intrinsic electronic properties of MHPs~\cite{Xiong2025,Li2024a,Miao2024} including Sn-based MHPs (CsSnX$_3$)~\cite{Liu2024a}. This approach eliminates reliance on extrapolated or approximate $n$, $k$ values.

But DFT cannot model the electromagnetic interactions that govern light propagation and extraction within PeLEDs~\cite{Rahimi2024}. So, to find that, DFT-driven electromagnetic modeling is required. The finite-difference time-domain (FDTD) method is a powerful approach that numerically solves Maxwell's equations in the time domain~\cite{Tabibifar2024}. This enables quantitative analysis of field distributions, resonant cavity effects, and photon outcoupling pathways~\cite{Cho2024}. Using DFT-derived $n$ and $k$ values as inputs, FDTD simulations can be used to calculate key performance metrics of PeLEDs. These include the Purcell factor, LEE, radiated power, spectral overlap, and far-field emission profiles~\cite{Rahimi2024,Tabibifar2024}.

Plasmonic coupling by embedding metallic nanoparticles is an effective strategy to improve light extraction in PeLEDs~\cite{Zhao2023}. This helps to overcome intrinsic optical limitations. Noble metal nanoparticles localize electromagnetic energy via surface plasmon resonance to enhance emission in PeLEDs~\cite{Bueno2025}. Among different nanoparticle geometries, Au nanorods (NR) offer superior chemical stability and oxidation resistance compared to Ag, along with enhanced local density of optical states (LDOS) that improves radiative recombination~\cite{Gu2020}. The longitudinal plasmon resonance of Au NRs can be tuned via aspect ratio control to align with NIR emission wavelengths~\cite{Ain2024}, making them particularly suitable for enhancing NIR-emitting CsSn$_x$Ge$_{1-x}$I$_3$ through optimized spectral overlap and Purcell enhancement. Since Au is biocompatible and resistant to environmental degradation, it is well suited for wearable and flexible optoelectronic applications~\cite{Li2025}.

Despite recent progress in lead-free perovskites, the relationship between Ge substitution and the optical properties of CsSn$_x$Ge$_{1-x}$I$_3$ PeLEDs has not been systematically investigated. Previous studies have focused primarily on stability improvements and bandgap tuning for photovoltaics, leaving the impact of Ge alloying on LED emission profiles, plasmonic interactions, and far-field characteristics largely unexplored.

We address this gap through an integrated DFT-FDTD framework that, for the first time, establishes quantitative links between Ge substitution, composition-specific optical properties, and device-level light extraction performance in CsSn$_x$Ge$_{1-x}$I$_3$ PeLEDs. We compute wavelength-dependent $n$ and $k$ values directly from DFT for each composition, enabling physically accurate electromagnetic modeling of emission enhancement, plasmonic coupling with Au NRs, and far-field radiation characteristics. We also optimize NR geometries across $x = 0$ to $1$ to achieve high LEE, Purcell enhancement, and spectral overlap. This makes NIR emission in device architectures compatible with flexible, wearable applications.

This paper is organized into four sections. Section 2 discusses computational methods for material and device analysis. Results and discussion are presented in Section 3, and Section 4 concludes the work.

\begin{figure}[htbp]
\centering\includegraphics[width=\linewidth]{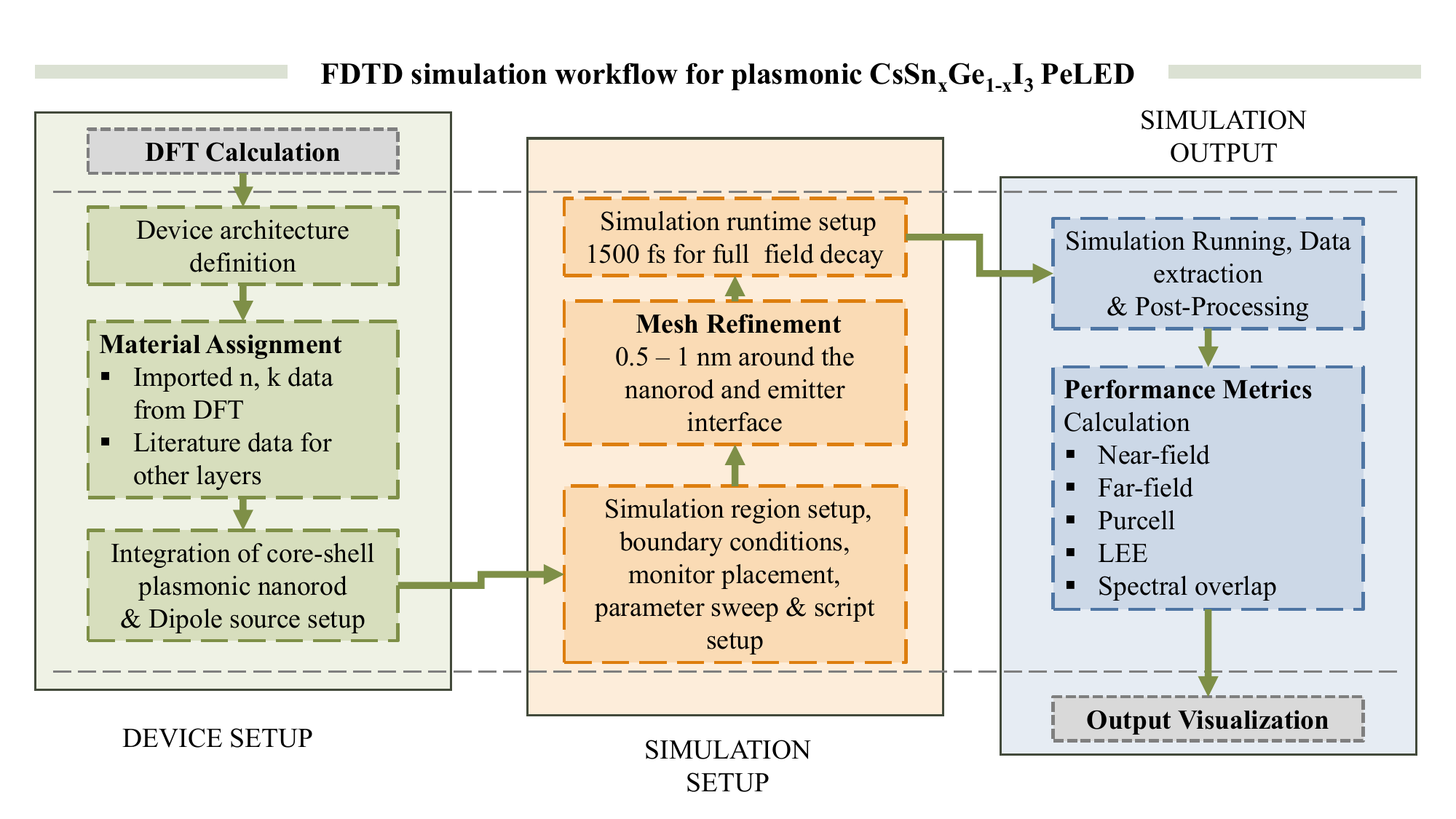}
\caption{Computational workflow for FDTD simulation of plasmonic CsSn$_x$Ge$_{1-x}$I$_3$-based PeLEDs. Device setup incorporates DFT-derived optical constants for the perovskite layer alongside literature data for other materials. FDTD simulations employ refined meshing near critical interfaces and integrate Au/SiO$_2$ NRs at the ZnO/perovskite boundary. Performance analysis yields near-field and far-field distributions, Purcell factors, light extraction efficiency, and spectral overlap metrics, enabling composition-dependent device optimization.}
\label{fig:workflow}
\end{figure}

\section{Computational Methods for Material and Device Analysis}

Our computational approach connects atomic-scale physics with device-scale optics by combining DFT and FDTD simulations. DFT provides the composition-dependent refractive index ($n$) and extinction coefficient ($k$) of CsSn$_x$Ge$_{1-x}$I$_3$ alloys, capturing their underlying electronic structure. These optical constants are then used as inputs for FDTD simulations of the complete PeLED architecture, enabling quantitative analysis of light emission, plasmonic enhancement, and spectral overlap with compositional accuracy.

\begin{figure}[htbp]
    \centering

    \begin{minipage}[b]{0.8\linewidth}
        \centering
        \includegraphics[width=\linewidth]{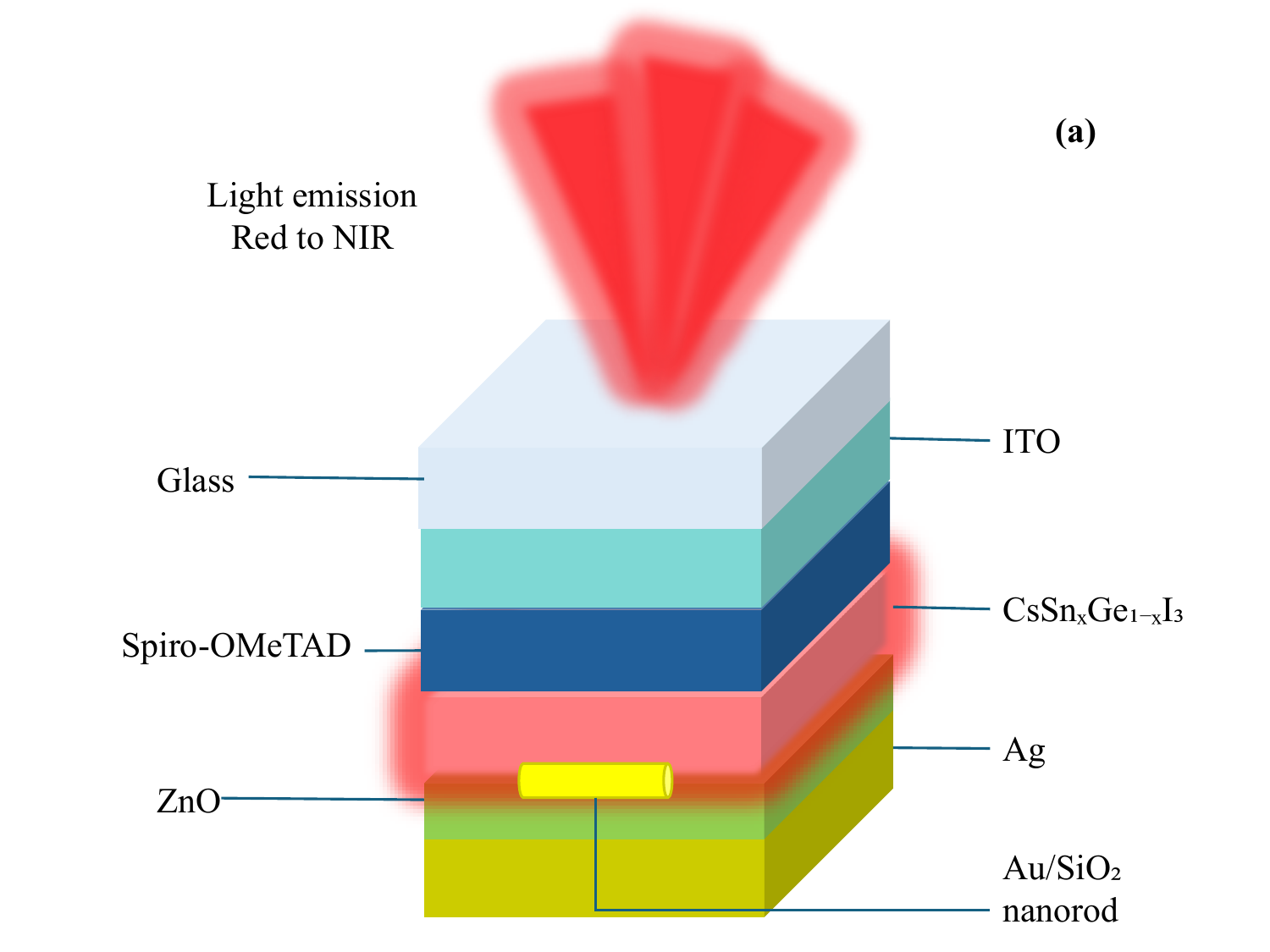}
    \end{minipage}
    \hfill
    \begin{minipage}[b]{0.9\linewidth}
        \centering
        \includegraphics[width=\linewidth]{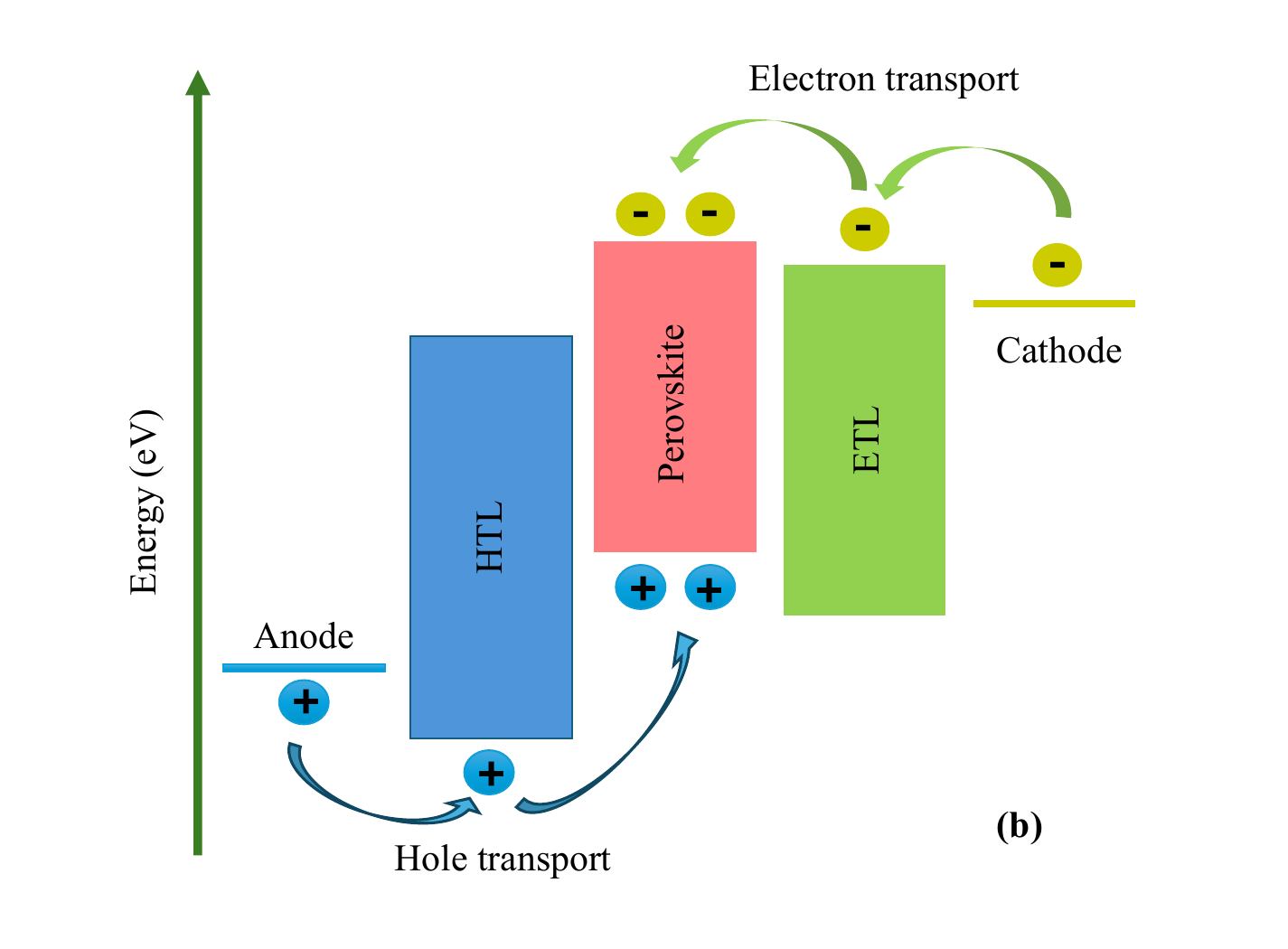}
    \end{minipage}

    \caption{(a) Layered device architecture of the CsSn$_x$Ge$_{1-x}$I$_3$-based PeLED, illustrating the anode/HTL/perovskite/ETL/cathode stack and the upward light-emission direction. (b) Energy-band diagram of the ITO/Spiro-OMeTAD/CsSn$_x$Ge$_{1-x}$I$_3$/ZnO/Ag structure, showing hole injection from ITO into Spiro-OMeTAD, electron injection from Ag into ZnO, and carrier recombination within the perovskite layer.}
    \label{fig:device}
\end{figure}

\subsection{First-Principles of Material Properties}

Density functional theory (DFT) calculations were performed using the CASTEP module in Materials Studio 2020. The study utilized the Generalized Gradient Approximation (GGA) with the Perdew-Burke-Ernzerhof (PBE) exchange-correlation functional. All calculations employed an ultrasoft pseudopotential with a plane-wave basis set and cutoff energy of 500~eV. For the Brillouin zone sampling, a Monkhorst--Pack k-point grid of ($6 \times 6 \times 6$) was used. The geometry of CsSn$_x$Ge$_{1-x}$I$_3$ was relaxed until the energy and force reached the criteria of convergence at $1.0 \times 10^{-6}$~eV and 0.01~eV~\AA$^{-1}$ between two adjacent steps. 

\subsection{Device Level Optical Modeling of the Perovskites}

3D FDTD simulations were performed using Lumerical FDTD (Ansys Inc.) to investigate the optical response of plasmonic CsSn$_x$Ge$_{1-x}$I$_3$-based PeLEDs. The FDTD algorithm solves Maxwell's equations numerically in the time domain, allowing modeling of light propagation, near-field enhancement, and resonant-cavity effects in multilayer optoelectronic structures~\cite{Taflove2005}. The computational workflow for FDTD simulation is shown in Fig.~\ref{fig:workflow}.

\begin{figure}[htbp]
    \centering
    \includegraphics[width=\linewidth]{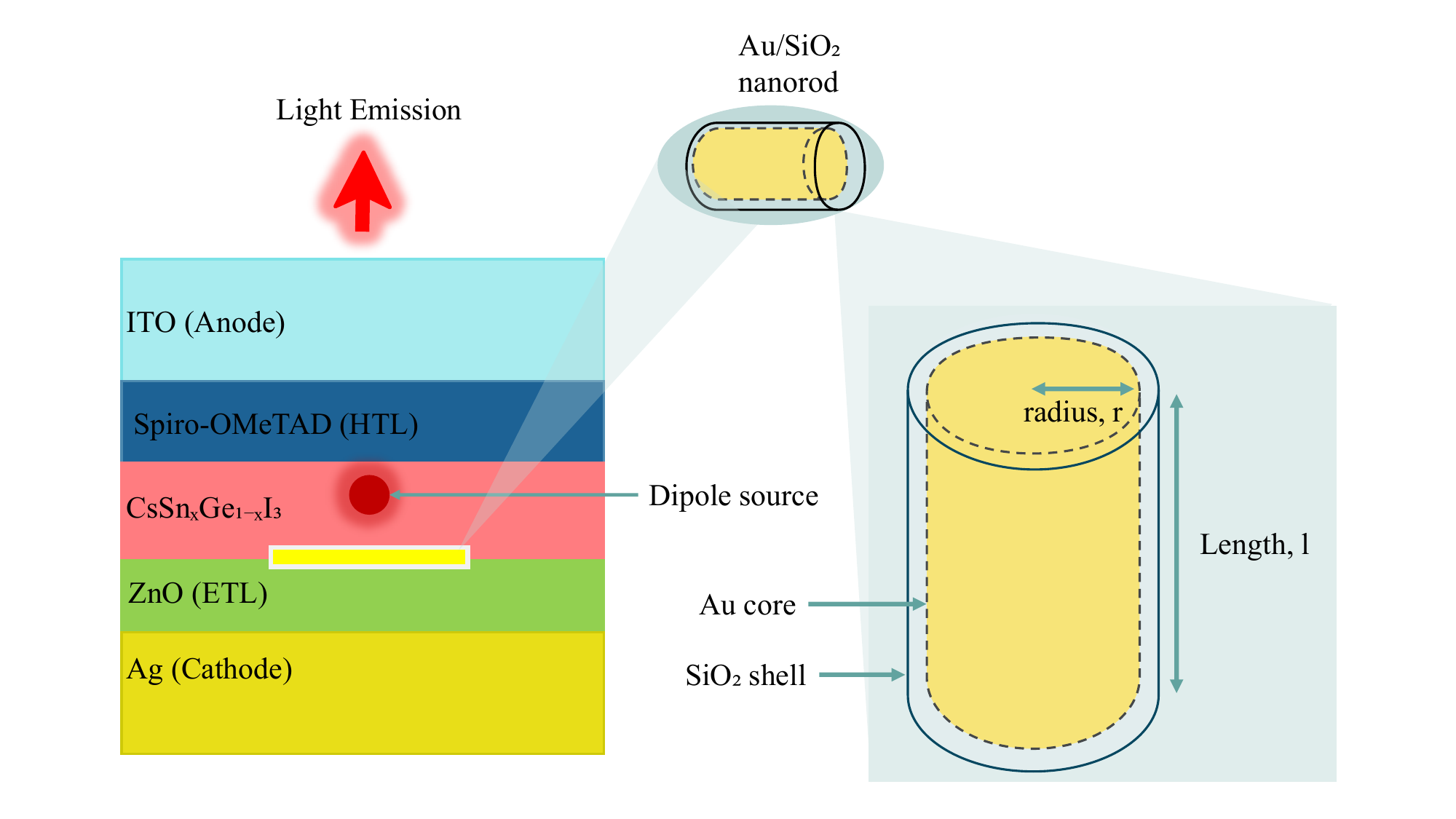}
    \caption{Schematic illustration of the CsSn$_x$Ge$_{1-x}$I$_3$-based PeLED structure incorporating a plasmonic Au/SiO$_2$ nanorod. The device stack (ITO/Spiro-OMeTAD/CsSn$_x$Ge$_{1-x}$I$_3$/ZnO/Ag) includes a dipole emitter positioned near the embedded Au/SiO$_2$ core--shell nanorod to enable plasmon--emitter coupling. The nanorod geometry is defined by the Au core length ($l$) and radius ($r$), surrounded by a SiO$_2$ shell.}
    \label{fig:nanorod}
\end{figure}

The simulated PeLED adopted an architecture of ITO (100~nm) / Spiro-OMeTAD (35~nm) / CsSn$_x$Ge$_{1-x}$I$_3$ (50~nm) / ZnO (40~nm) / Ag (100~nm), matching experimentally achievable stacks (Fig.~\ref{fig:device}a). Fig.~\ref{fig:device}b represents the energy-band diagram of the ITO /Spiro-OMeTAD /CsSn$_x$Ge$_{1-x}$I$_3$ /ZnO /Ag structure, showing hole injection from ITO into Spiro-OMeTAD, electron injection from Ag into ZnO, and carrier recombination within the perovskite layer. Composition-dependent $n$, $k$ spectra for the perovskite layer were derived from DFT calculations and imported into the FDTD material database. Optical constants for the remaining layers were taken from established datasets~\cite{Brivio2014,Palik1985,Alnuaimi2016,Johnson1972}. Spontaneous emission within the perovskite layer was represented by an electric dipole source positioned above the plasmonic nanostructure, radiating across the visible and NIR wavelengths corresponding to the emission window of CsSn$_x$Ge$_{1-x}$I$_3$ compositions (Fig.~\ref{fig:nanorod}). This source model treats electron-hole radiative recombination as an oscillating dipole and reproduces Purcell-enhanced emission in resonant photonic environments~\cite{Purcell1946}. For plasmonic enhancements, a gold core-shell structured NR (Fig.~\ref{fig:nanorod}) was positioned near the ZnO/perovskite interface, with SiO$_2$ acting as the shell to protect charge carriers from quenching~\cite{Cui2024MEFHgSensor}. The NR diameter and length were tuned so that the localized surface plasmon resonance (LSPR) spectrally overlapped with the composition-dependent emission peak. This produced strong near-field coupling and enhanced radiative decay~\cite{Maier2007,Lozano2013}. The frequency-dependent dielectric function of gold was defined according to the Palik dataset~\cite{Palik1985}.

Periodic boundary conditions were applied in the lateral ($x$--$y$) directions to emulate an infinite emitter array, while perfectly matched layer (PML) boundaries were implemented vertically ($z$) to suppress spurious reflections~\cite{Berenger1994}. This configuration reproduces realistic outcoupling behavior and has been validated for photonic-band-structure and plasmonic-resonator analyses~\cite{Montano-Priede2022}. A non-uniform spatial mesh was adopted, with refinement to 0.5--1~nm around the NR and emitter interface, gradually coarsening elsewhere to maintain computational efficiency. Each simulation was executed for 1500~fs, ensuring full electromagnetic-field decay before Fourier transformation. From the computed field data, several key optical metrics were extracted.

The resulting far-field emission patterns and optical enhancement metrics are analyzed in Section 3 to establish composition-structure-performance relationships.

\section{Results and Discussion}

\subsection{Electrical and Optical Properties of CsSn$_x$Ge$_{1-x}$I$_3$}

\begin{figure}[htbp]
\centering\includegraphics[width=0.9\linewidth]{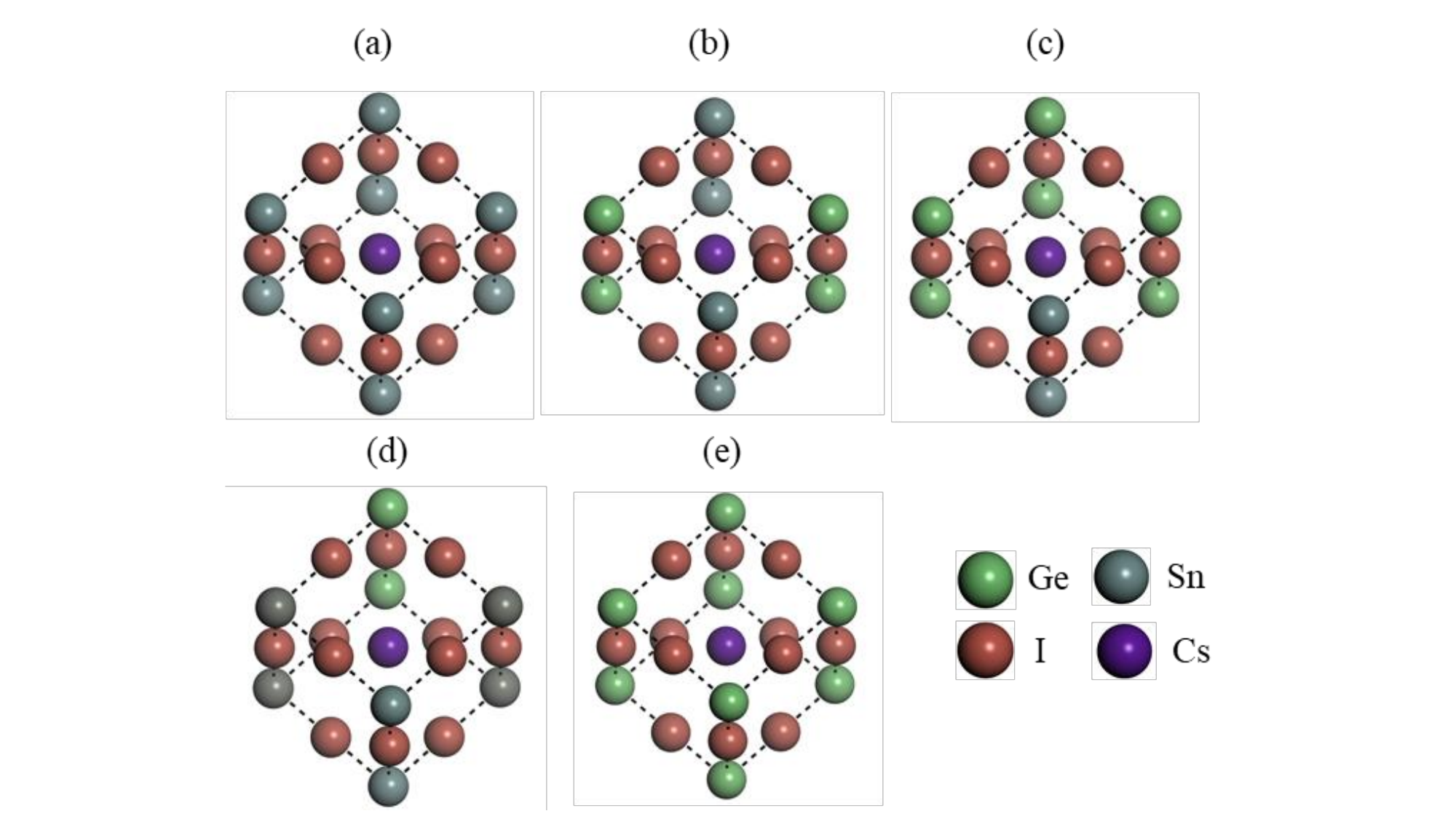}
\caption{Different doping arrangement of CsSn$_x$Ge$_{1-x}$I$_3$. (a) $x=1$, (b) $x=0.75$, (c) $x=0.5$, (d) $x=0.25$ and (e) $x=0$. Green, grey, brown, and violet indicate Ge, Sn, I, and Cs, respectively.}
\label{fig:doping}
\end{figure}

The structural, electronic, and optical properties of CsSn$_x$Ge$_{1-x}$I$_3$ ($0 \leq x \leq 1$) were systematically investigated using  density functional theory (DFT) to establish the composition-property relationships governing light emission and plasmonic interactions in PeLED architectures. Five representative alloy configurations are pure CsSnI$_3$ ($x=1$), CsSn$_{0.75}$Ge$_{0.25}$I$_3$, CsSn$_{0.5}$Ge$_{0.5}$I$_3$, CsSn$_{0.25}$Ge$_{0.75}$I$_3$, and pure CsGeI$_3$ ($x=0$) respectively, were analyzed to understand the effect of doping gradually (Fig.~\ref{fig:doping}). All structures converged to the orthorhombic perovskite phase, maintaining corner-sharing octahedra structure. The lattice volume decreased monotonically with increasing Ge content, consistent with the smaller ionic radius of Ge$^{2+}$ relative to Sn$^{2+}$. This contraction leads to stronger metal-halide orbital overlap and influences both band dispersion and optical transition strength.

The absorption coefficient can be derived from the complex dielectric constant, $\epsilon = \epsilon_1 + i\epsilon_2$. The imaginary part ($\epsilon_2$) of the dielectric constant ($\epsilon$) can be calculated using the following formula from the electrical band structure~\cite{Gajdos2006}:
\begin{equation}
\epsilon_2\left(q \rightarrow 0_{\hat{u}}, \hbar\omega\right) = \frac{2e^2\pi}{\Omega\epsilon_0} \sum_{k,v,c} \left|\langle\Psi_k^c|\mathbf{u} \cdot \mathbf{r}|\Psi_k^v\rangle\right|^2 \delta\left(E_k^c - E_k^v - \hbar\omega\right)
\label{eq:epsilon2}
\end{equation}
where $\mathbf{u}$ is the vector defining the polarization of the incident light, $\langle\Psi_k^c|\mathbf{u} \cdot \mathbf{r}|\Psi_k^v\rangle$ is the matrix element, $E_k^c$ is the conduction band energy, $E_k^v$ is the valence band energy at wave number $k$, $\omega$ is the angular frequency of the electron, $e$ is the charge, $\mathbf{u} \cdot \mathbf{r}$ is the momentum operator, and $\hbar$ is the reduced Planck's constant. The real part of the dielectric constant can be found using the Kramers--Kronig transformation~\cite{Palik1998}:
\begin{equation}
\epsilon_1(\omega) = 1 + \frac{2}{\pi} P \int_0^{\infty} \frac{\epsilon_2(\omega^*)\omega^*}{\omega^{*2} - \omega^2} d\omega^*
\label{eq:epsilon1}
\end{equation}
where $P$ is the principal value of the integral. The optical absorption coefficient for the material is calculated as a function of dielectric constants as follows~\cite{Wan2013}:
\begin{equation}
\alpha(\omega) = \frac{4\pi\kappa(\omega)}{\lambda\sqrt{2}} \left(\sqrt{\epsilon_1^2(\omega) + \epsilon_2^2(\omega)} - \epsilon_1(\omega)\right)^{1/2}
\label{eq:absorption}
\end{equation}
where $\epsilon_1$ and $\epsilon_2$ are the real and imaginary parts of the dielectric function, $\kappa$ is the extinction coefficient, and $\lambda$ is the wavelength. The refractive index ($n$) and extinction coefficient ($k$) can be obtained from the dielectric constant by the following formulas:
\begin{equation}
n(\omega) = \frac{1}{\sqrt{2}} \left[\sqrt{\epsilon_1^2(\omega) + \epsilon_2^2(\omega)} + \epsilon_1(\omega)\right]^{1/2}
\label{eq:n}
\end{equation}
\begin{equation}
k(\omega) = \frac{1}{\sqrt{2}} \left[\sqrt{\epsilon_1^2(\omega) + \epsilon_2^2(\omega)} - \epsilon_1(\omega)\right]^{1/2}
\label{eq:k}
\end{equation}

\begin{figure}[htbp]
\centering\includegraphics[width=\linewidth]{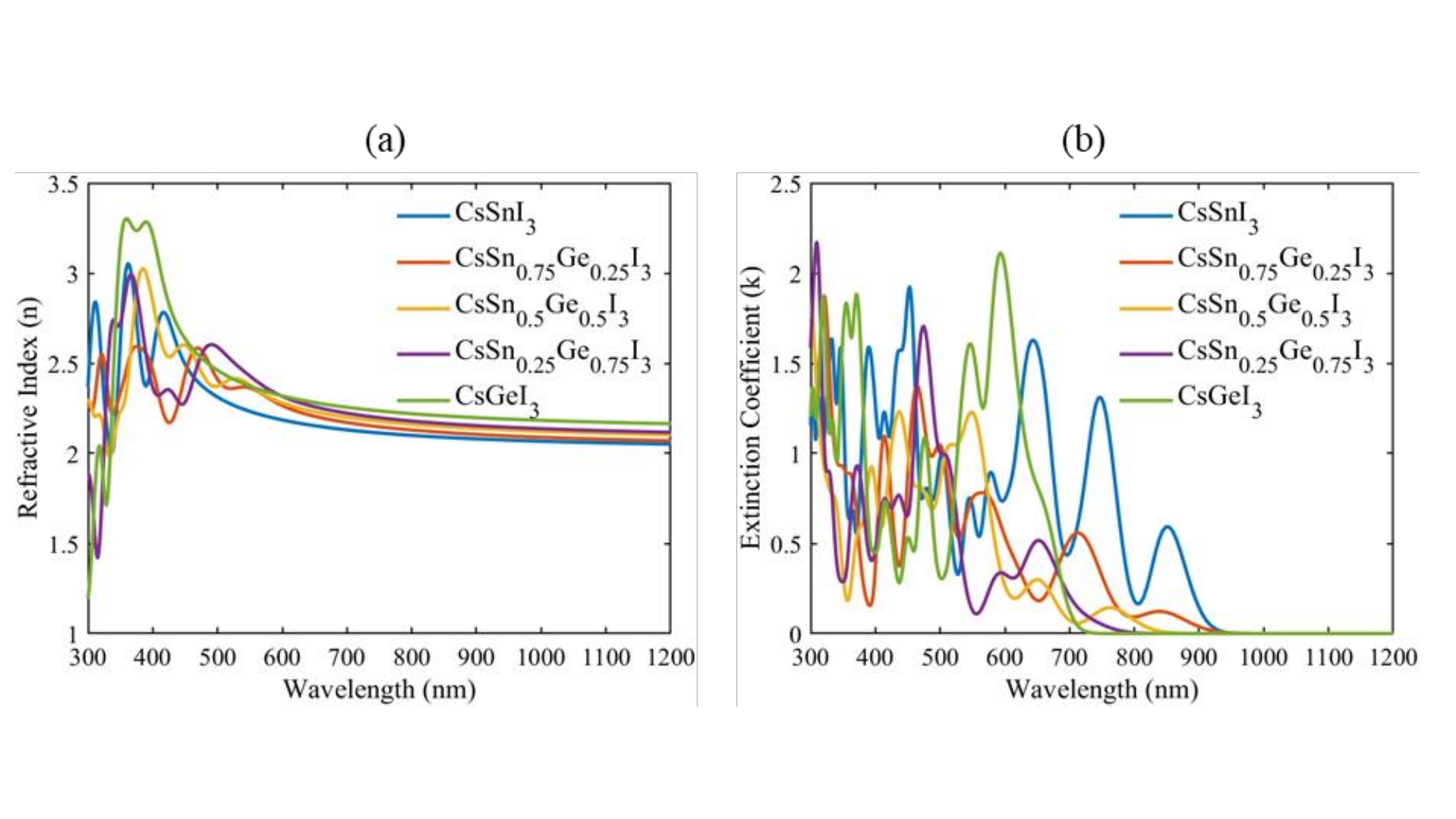}
\caption{Optical properties of CsSn$_x$Ge$_{1-x}$I$_3$. (a) Refractive index and (b) extinction coefficient for various compositions of $x$.}
\label{fig:optical}
\end{figure}

The complex refractive index ($n$ and $k$ spectra) obtained from the frequency-dependent dielectric function reveals pronounced compositional dependence (Fig.~\ref{fig:optical}). Across all compositions, the refractive index exhibits a strong peak near the excitonic transition region, gradually decreasing at higher photon energies. Increasing Ge substitution systematically blue shifts the $n(\lambda)$ and $k(\lambda)$ spectra, reflecting the widening electronic bandgap. Importantly, the magnitude of $n$ remains high ($\sim$2.2--2.6 in the visible/NIR region), corroborating the inherently strong light-confinement characteristics of Sn/Ge-based perovskites and explaining the limited intrinsic light extraction efficiency in planar PeLEDs.

\begin{figure}[htbp]
\centering\includegraphics[width=\linewidth]{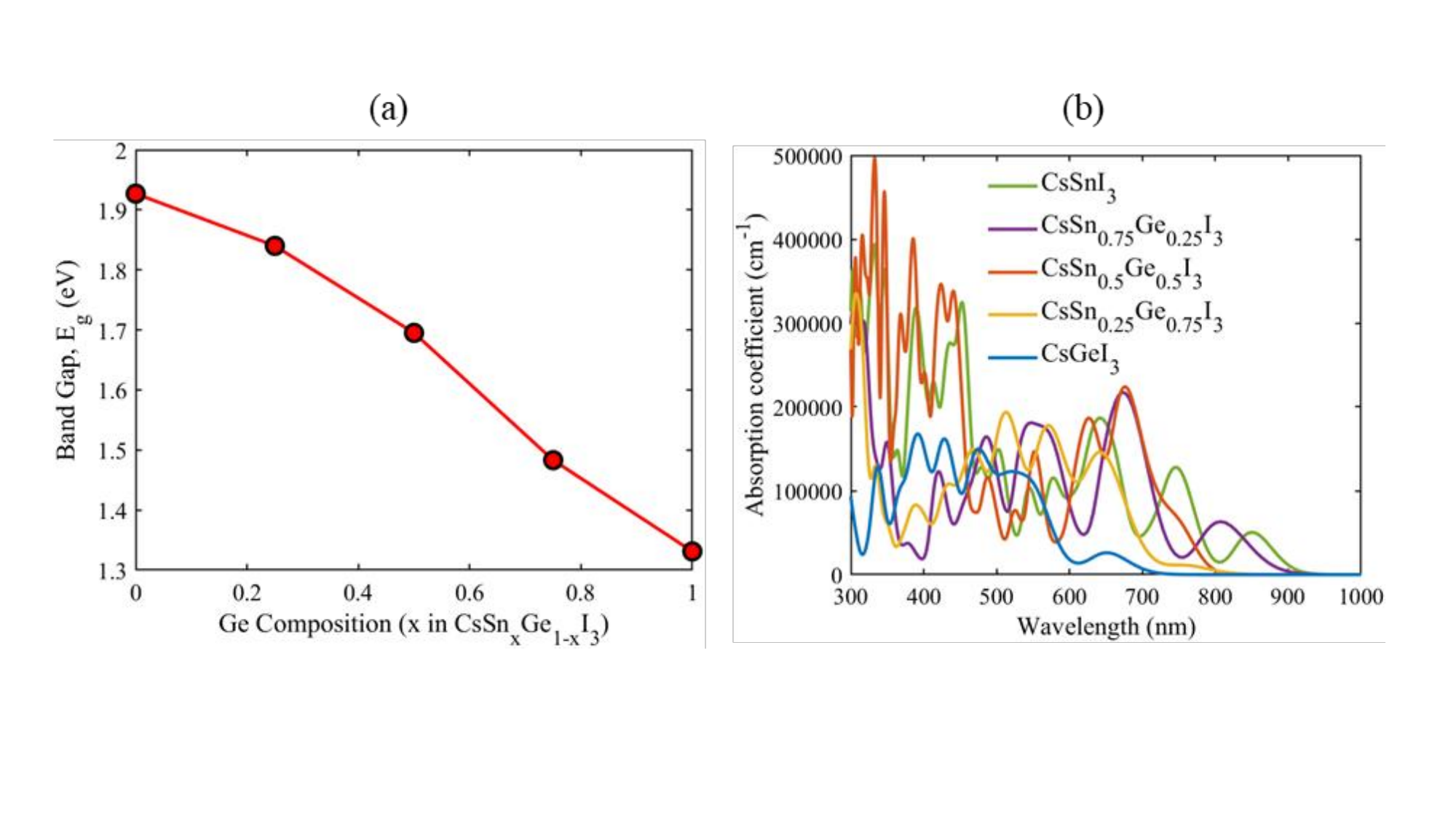}
\caption{(a) Bandgap of CsSn$_x$Ge$_{1-x}$I$_3$ for various compositions of $x$. (b) Absorption coefficient of CsSn$_x$Ge$_{1-x}$I$_3$ for various compositions of $x$.}
\label{fig:bandgap}
\end{figure}

The absorption coefficients in Fig.~\ref{fig:bandgap}(b) further confirm the influence of Ge incorporation on optical transitions. Pure CsSnI$_3$ exhibits strong absorption extending into the NIR, whereas Ge-rich compositions shift the absorption edge toward shorter wavelengths, consistent with increased bandgap energies. The absorption amplitude remains high ($>10^5$~cm$^{-1}$ across the visible--NIR window), reaffirming the exceptional oscillator strength and suitability of these materials for high-radiance emitter applications. Alloy compositions ($x=0.25$--$0.75$) show intermediate absorption edges, enabling fine-tuning of emission wavelengths for targeted plasmonic coupling with Au nanorods.

The electronic bandgaps analysis of the DFT optical spectra displays a nearly linear dependence on alloy composition (Fig.~\ref{fig:bandgap}a). CsSnI$_3$ shows the smallest bandgap of 1.331~eV, while CsGeI$_3$ exhibits a significantly larger value of 1.927~eV, with mixed alloys occupying the intermediate range. Ge substitution reduces the antibonding interaction strength in the metal--iodide network, widening the bandgap and shifting emission toward shorter wavelengths. Such tunability is crucial because it permits precise matching of the emissive peak with the longitudinal plasmon resonance of Au/SiO$_2$ nanorods for maximized Purcell enhancement.

The DFT results establish a clear composition-dependent evolution of lattice structure, bandgap, and optical constants in CsSn$_x$Ge$_{1-x}$I$_3$. These properties constitute the fundamental inputs for the subsequent FDTD modeling.

\begin{table}[htbp]
\caption{FDTD Results}
\label{tab:fdtd}
\centering
\footnotesize
\setlength{\tabcolsep}{2.5pt}
\renewcommand{\arraystretch}{1.15}

\begin{tabularx}{\linewidth}{c Y c c c c c c c c}
\toprule
\multirow{2}{*}{$x$} & \multirow{2}{*}{Composition} & Emission & NR Length & NR Radius &
\multicolumn{2}{c}{Purcell} & LEE & LEE Enh. & Sp. Ovlp. \\
 &  & $\lambda$ (nm) & (nm) & (nm) & Peak & $\lambda$ (nm) & (\%) & (\%) & $J_{\cos}$ \\
\midrule
0 & CsGeI$_3$ & 643 & 55 & 11 & 4.4 & 643 & 24.9 & 17 & 0.89 \\
0.25 & CsSn$_{0.25}$Ge$_{0.75}$I$_3$ & 674 & 60 & 11 & 12.1 & 695 & 19 & 34 & 0.80 \\
0.5 & CsSn$_{0.5}$Ge$_{0.5}$I$_3$ & 731 & 70 & 17 & 5.3 & 721 & 25 & 33 & 0.93 \\
0.75 & CsSn$_{0.75}$Ge$_{0.25}$I$_3$ & 836 & 120 & 17 & 4.9 & 837 & 23 & 28 & 0.96 \\
1 & CsSnI$_3$ & 931 & 170 & 19 & 8.0 & 928 & 17.5 & 36 & 0.96 \\
\bottomrule
\end{tabularx}
\end{table}

\subsection{FDTD Results}

\begin{figure}[htbp]
\centering

\begin{subfigure}{0.48\textwidth}
  \centering
  \caption*{(a)}
  \includegraphics[width=\linewidth]{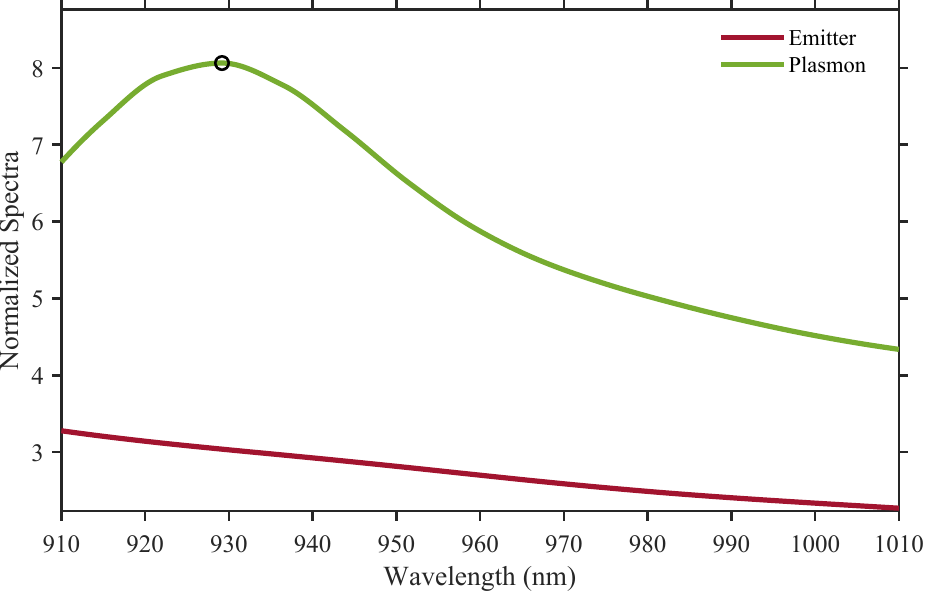}
\end{subfigure}
\hfill
\begin{subfigure}{0.48\textwidth}
  \centering
  \caption*{(b)}
  \includegraphics[width=\linewidth]{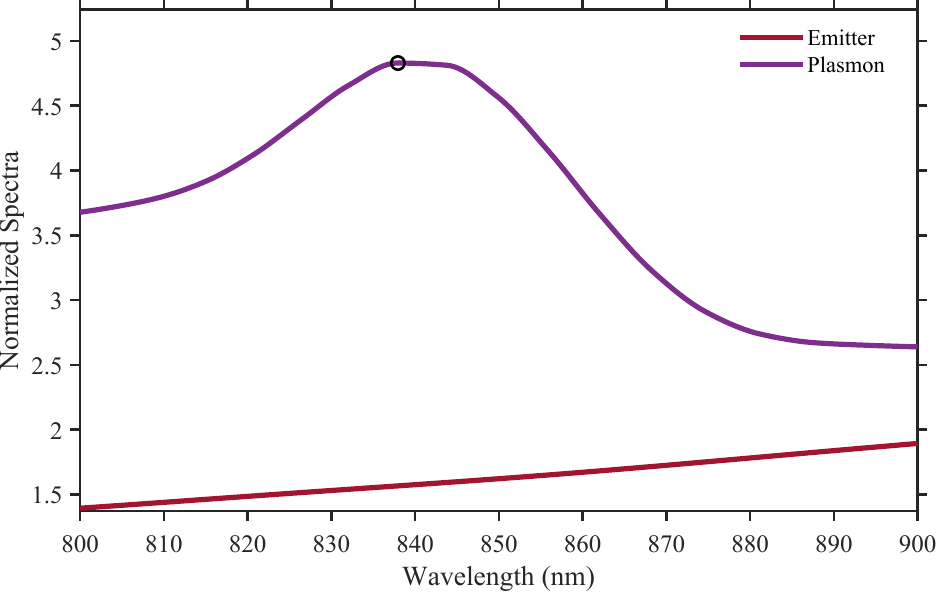}
\end{subfigure}

%\vspace{3mm}

\begin{subfigure}{0.48\textwidth}
  \centering
  \caption*{(c)}
  \includegraphics[width=\linewidth]{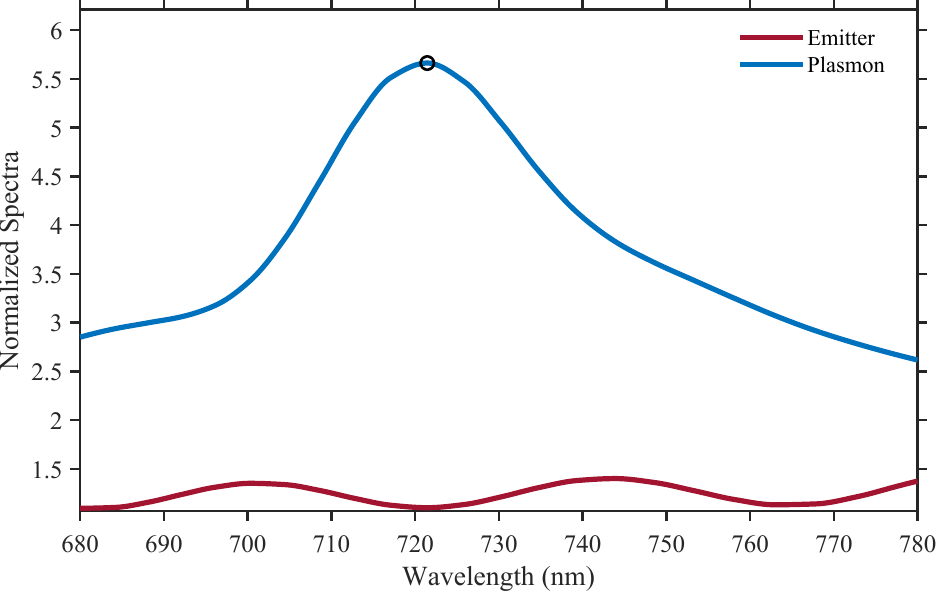}
\end{subfigure}
\hfill
\begin{subfigure}{0.48\textwidth}
  \centering
  \caption*{(d)}
  \includegraphics[width=\linewidth]{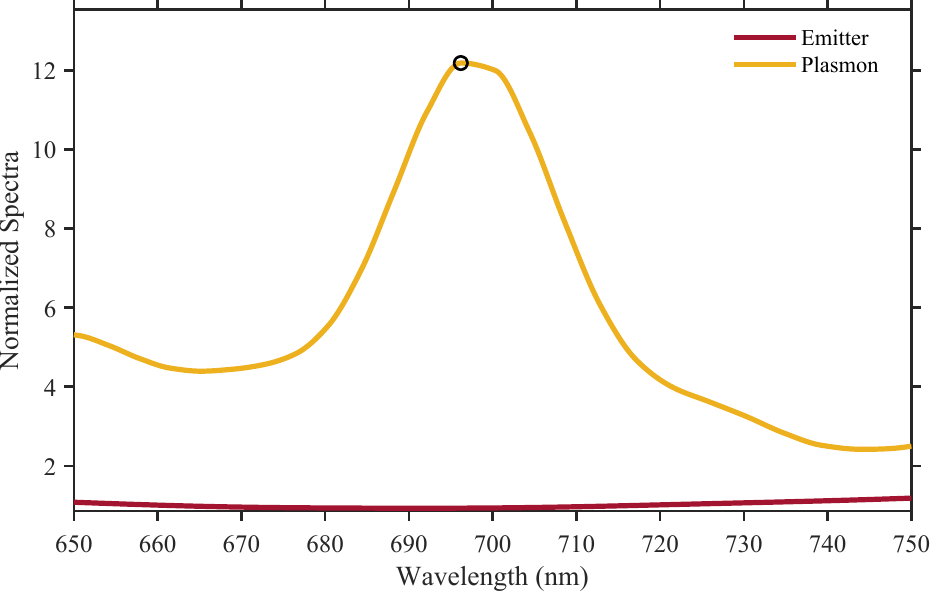}
\end{subfigure}

%\vspace{3mm}

\begin{subfigure}{0.48\textwidth}
  \centering
  \caption*{(e)}
  \includegraphics[width=\linewidth]{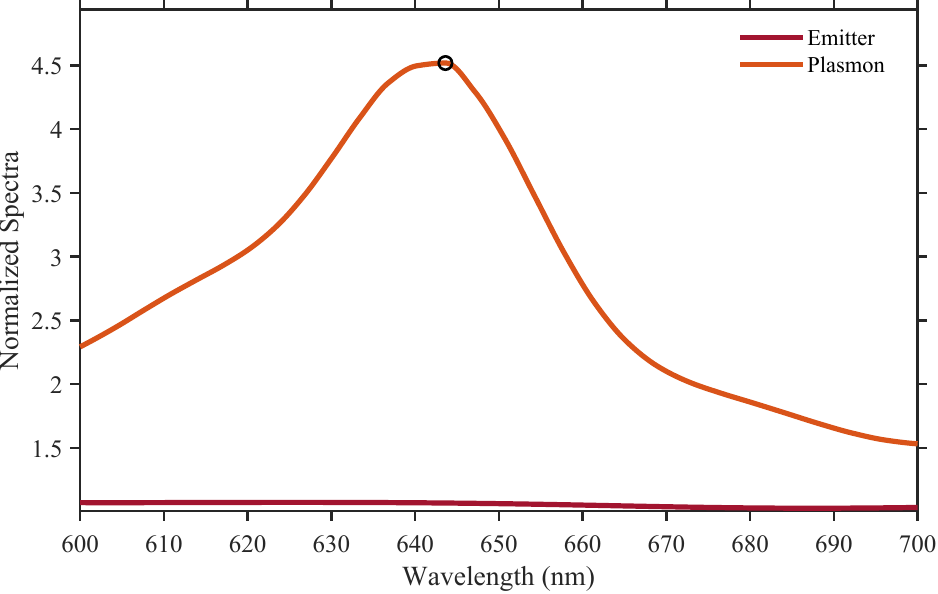}
\end{subfigure}

\caption{Composition-dependent Purcell factor enhancement in CsSn$_x$Ge$_{1-x}$I$_3$-based LED for (a) $x=1$, (b) $x=0.75$, (c) $x=0.5$, (d) $x=0.25$, and (e) $x=0$.}
\label{fig:purcell}
\end{figure}

\subsubsection{Purcell Factor Enhancement}

Purcell factor is a measure of the enhancement of emitter’s spontaneous decay rate when nanostructures are placed nearby~\cite{Zambrana-Puyalto2015}. Enhancement occurs due to the modification of local photonic density of states (LDOS) at the emitter. The Purcell factor enhancement of the LED is shown in Fig.~\ref{fig:purcell} for all the values of x. Black markers indicate peak values of Purcell across all x values. The highest Purcell factor of 12 was achieved for CsSn$_{0.25}$Ge$_{0.75}$I$_3$ for NR length of 60 nm and radius of 11 nm. This indicates strong coupling between the emitter and the plasmonic nanostructure. Table~\ref{tab:fdtd} shows Purcell enhancement for all compositions CsSn$_x$Ge$_{1-x}$I$_3$ and the corresponding NR length and radius values.

Coincidence of plasmon spectrum peak with the emitter peak creates resonance, increasing LDOS at that wavelength. This resonant condition in turn enhances the radiative decay rate of the emitter resulting in large Purcell factor~\cite{Zhao2020}. Other conditions where the plasmonic response is weaker or not in coincidence with the emitter peak, LDOS may still be increased due to scattering~\cite{Vurgaftman2024}. This is responsible for Purcell enhancements that are not sharp. Measured Purcell behavior is a superposition of both resonance-driven enhancement where spectral overlap is significant and broadband off-resonant LDOS perturbation where overlap is negligible~\cite{Zhao2020}.

\begin{figure}[htbp]
\centering
\begin{subfigure}{0.48\textwidth}
  \centering
  \caption*{(a)}
  \includegraphics[width=\linewidth]{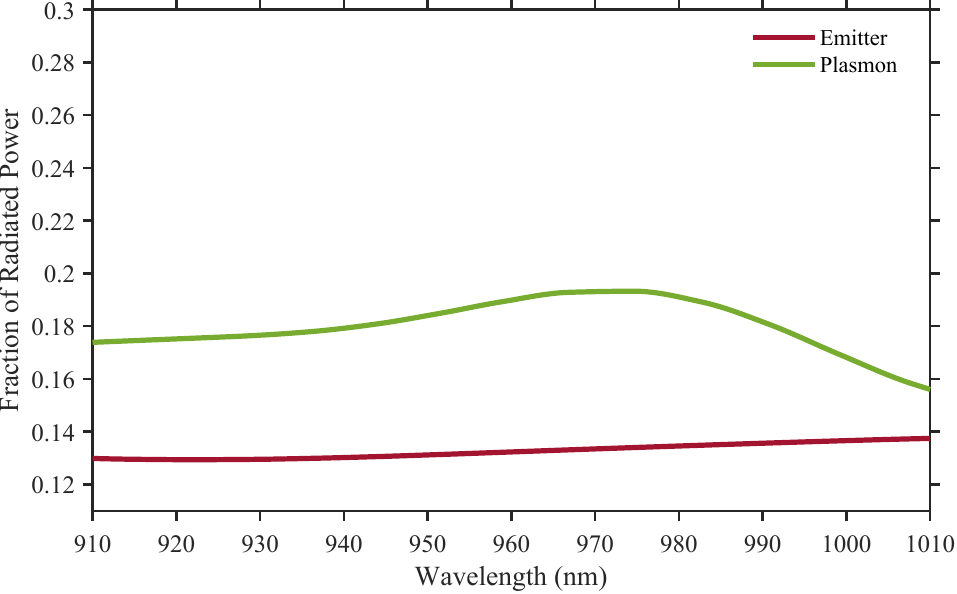}
\end{subfigure}
\hfill
\begin{subfigure}{0.48\textwidth}
  \centering
  \caption*{(b)}
  \includegraphics[width=\linewidth]{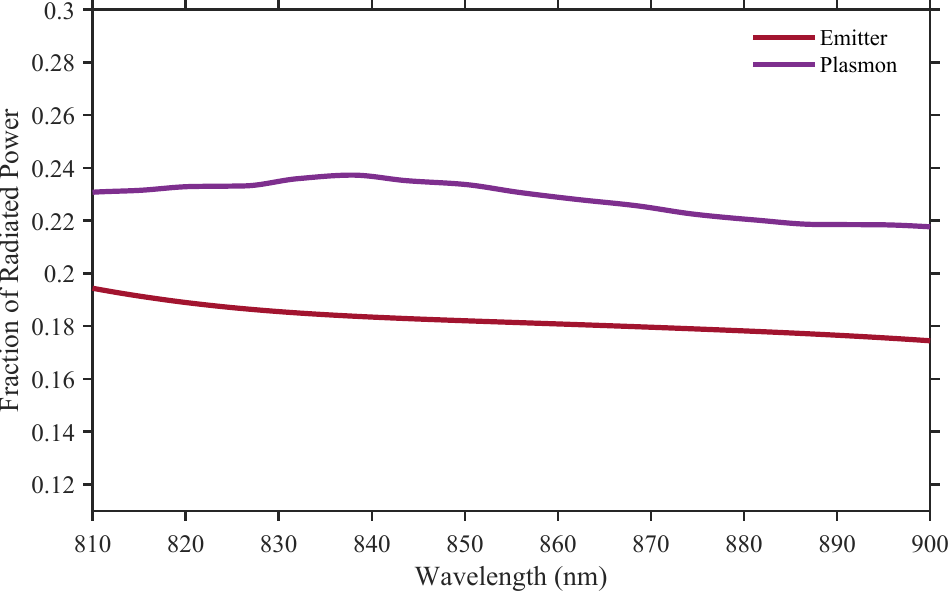}
\end{subfigure}
\begin{subfigure}{0.48\textwidth}
  \centering
  \caption*{(c)}
  \includegraphics[width=\linewidth]{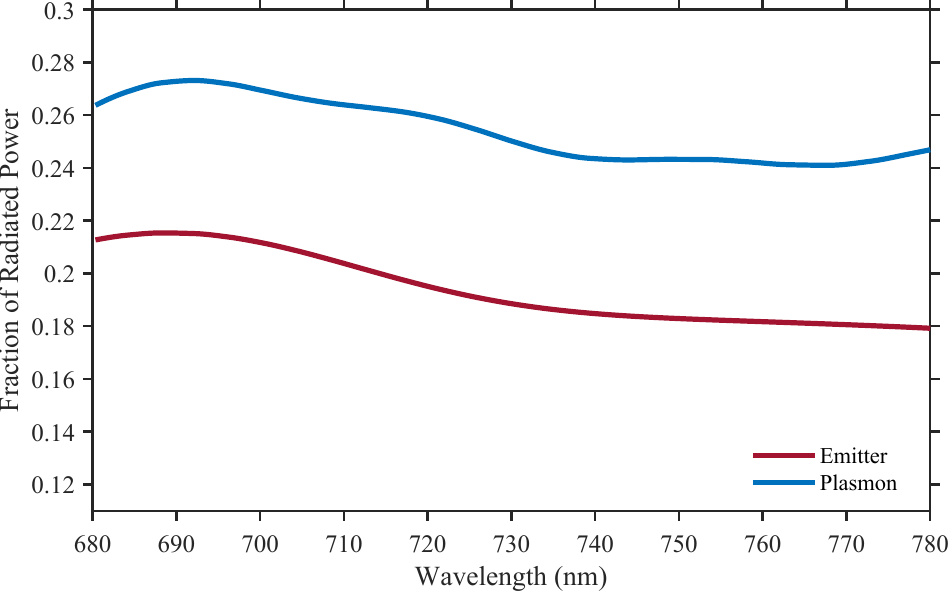}
\end{subfigure}
\hfill
\begin{subfigure}{0.48\textwidth}
  \centering
  \caption*{(d)}
  \includegraphics[width=\linewidth]{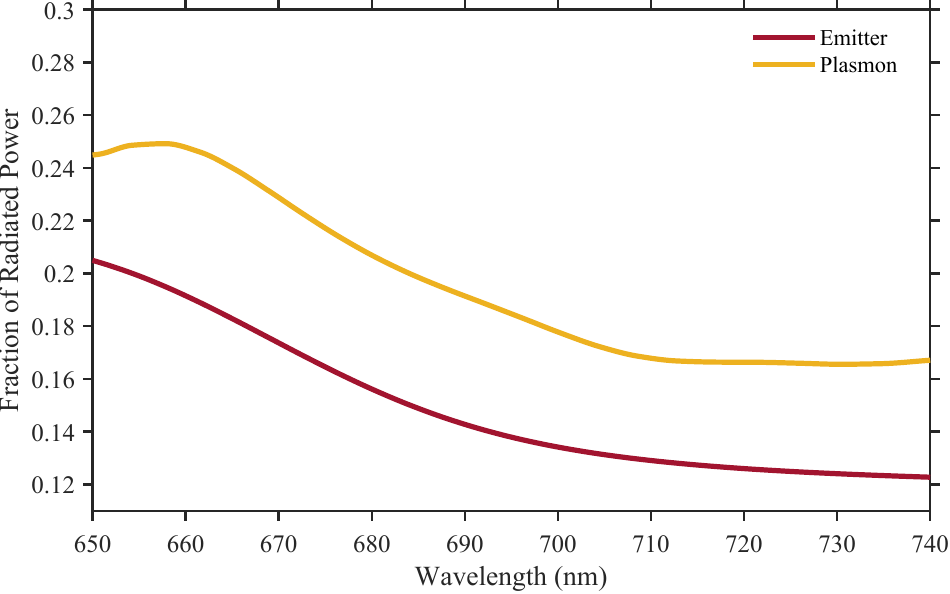}
\end{subfigure}
\begin{subfigure}{0.48\textwidth}
  \centering
  \caption*{(e)}
  \includegraphics[width=\linewidth]{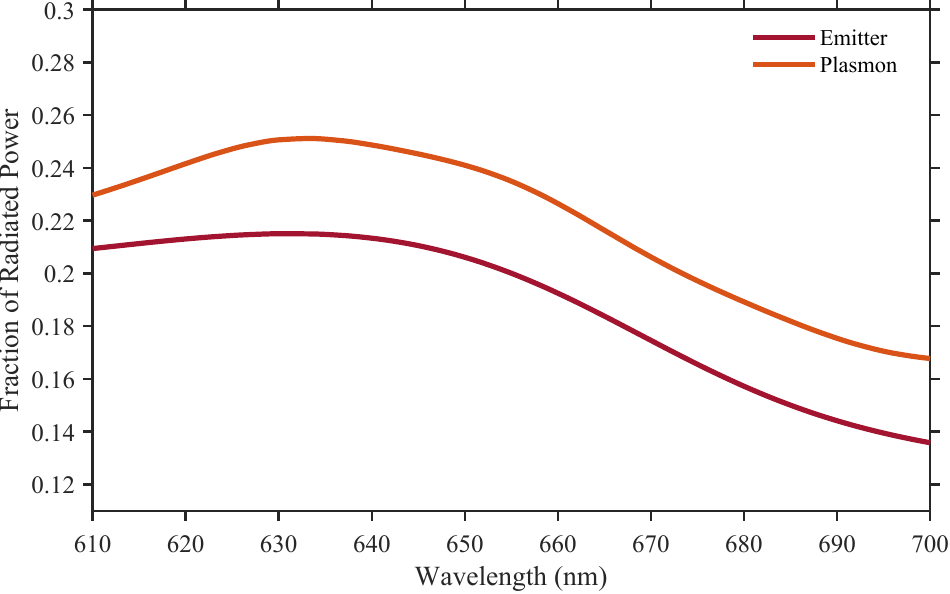}
\end{subfigure}
\caption{Composition-dependent LEE in CsSn$_x$Ge$_{1-x}$I$_3$-based LED for (a) $x = 1$, (b) $x =0.75$, (c) $x =0.5$, (d) $x =0.25$, (e) $x =0$.}
\label{fig:LEE}
\end{figure}

\subsubsection{LEE and LEE Enhancement}

\begin{figure}[htbp]
\centering
\begin{subfigure}{0.48\textwidth}
  \centering
  \caption*{(a)}
  \includegraphics[width=\linewidth]{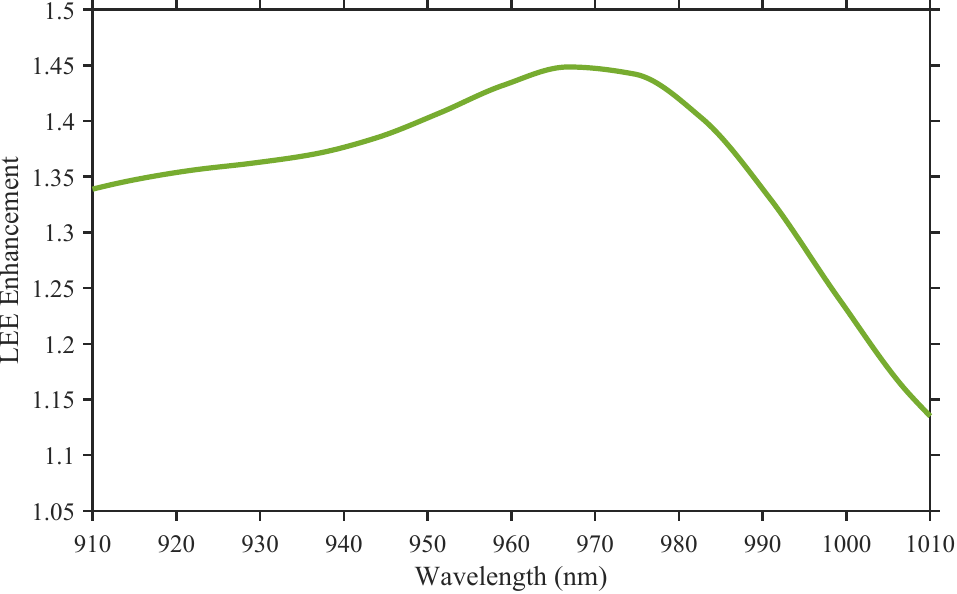}
\end{subfigure}
\hfill
\begin{subfigure}{0.48\textwidth}
  \centering
  \caption*{(b)}
  \includegraphics[width=\linewidth]{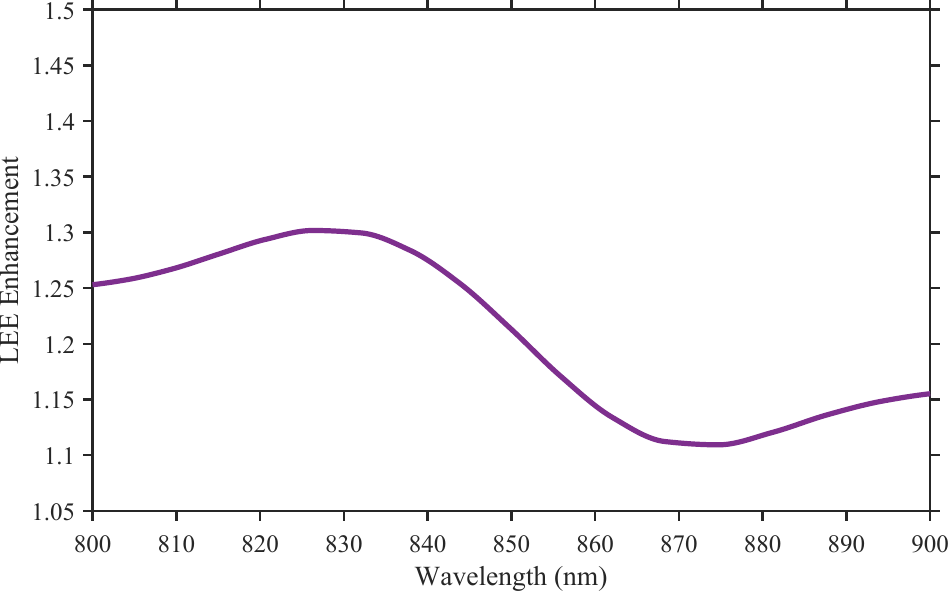}
\end{subfigure}
\begin{subfigure}{0.48\textwidth}
  \centering
  \caption*{(c)}
  \includegraphics[width=\linewidth]{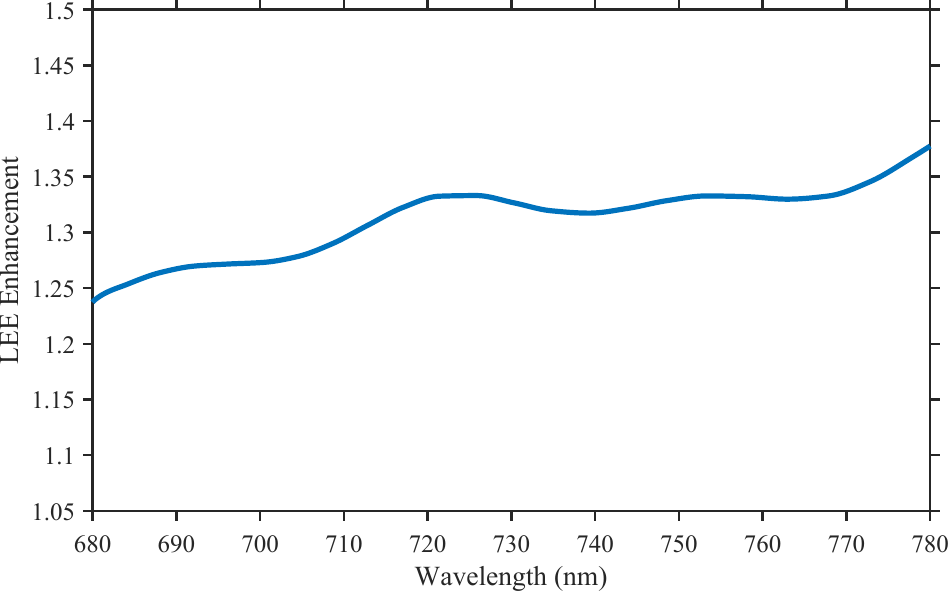}
\end{subfigure}
\hfill
\begin{subfigure}{0.48\textwidth}
  \centering
  \caption*{(d)}
  \includegraphics[width=\linewidth]{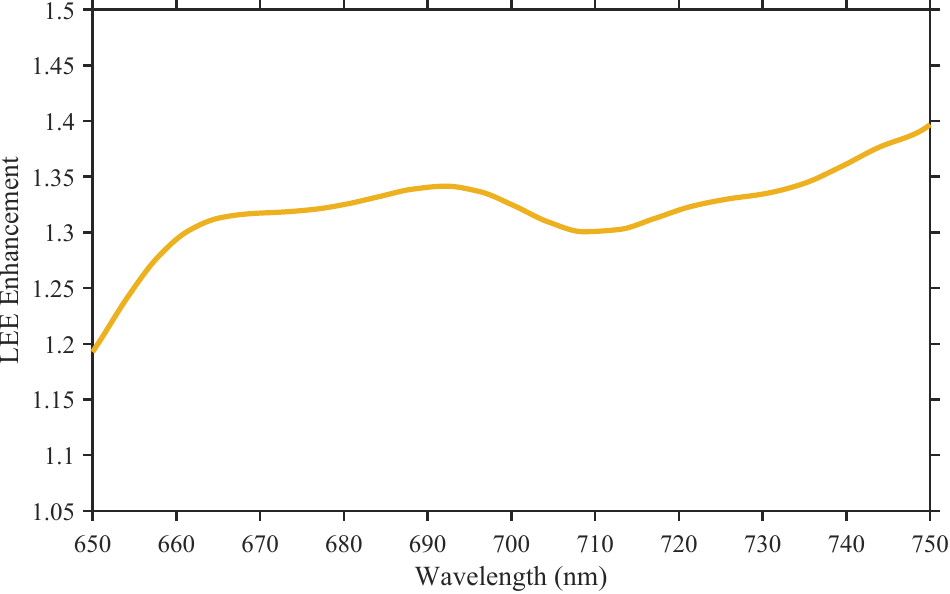}
\end{subfigure}
\begin{subfigure}{0.48\textwidth}
  \centering
  \caption*{(e)}
  \includegraphics[width=\linewidth]{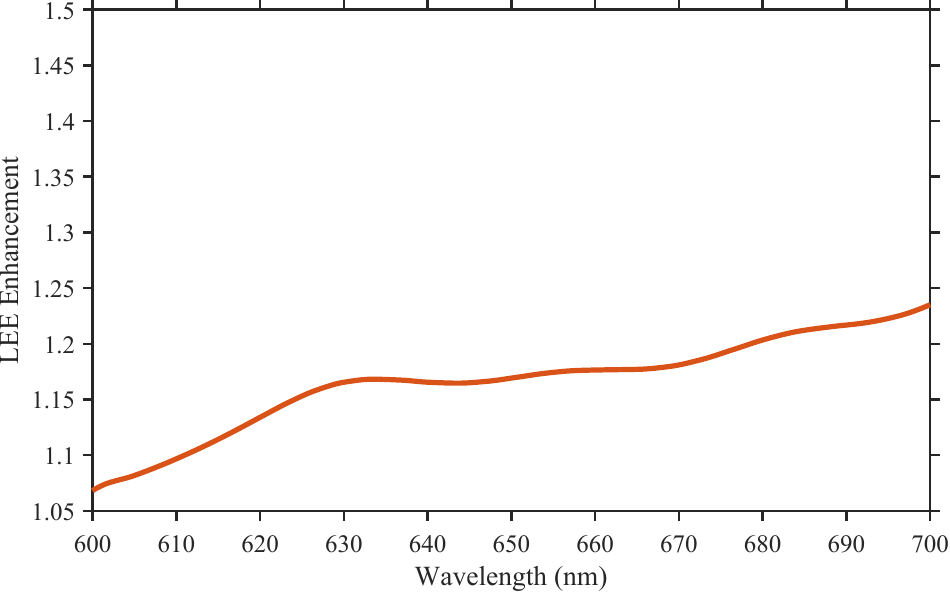}
\end{subfigure}
\caption{Composition-dependent LEE enhancement in CsSn$_x$Ge$_{1-x}$I$_3$-based LED for (a) $x = 1$, (b) $x =0.75$, (c) $x =0.5$, (d) $x =0.25$, (e) $x =0$.}
\label{fig:LEE_enhancement}
\end{figure}

The light extraction efficiency (LEE) quantifies the fraction of generated photons that escape the device structure~\cite{Raypah2022}. Without plasmonic nanorods, the baseline LEE for these high-index perovskites is approximately 8\%. Fig.~\ref{fig:LEE} and Fig.~\ref{fig:LEE_enhancement} show the composition-dependent LEE in CsSn$_x$Ge$_{1-x}$I$_3$-based LEDs. LEE can be seen consistently increasing with Ge content, reaching a maximum of 25\% for CsSn$_{0.5}$Ge$_{0.5}$I$_3$ using NR length of 70 nm and radius of 17 nm. The highest LEE enhancement of 36\% was obtained for CsSnI$_3$ ($x=1$) using NR length of 170 nm and radius of 19 nm. 

Several mechanisms work in synergy when Au NR is used for LEE and LEE enhancement. When NR is placed near the perovskite emitter, the nanorods support localized surface plasmon resonances (LSPR), opening additional radiative channels and enhancing the local density of optical states~\cite{Katyal2021}. The presence of NR breaks the symmetry of the device stack, scattering previously trapped waveguided modes into the far field and adding directionality to the emission~\cite{Ai2022}. This scattering contribution, combined with the Purcell-driven increase in radiative photon generation, elevates the fraction of photons that escape the structure~\cite{Katyal2021}. NR geometry also allows precise resonance tuning and large scattering cross-sections, further boosting both LEE and LEE enhancement across compositions.

Table~\ref{tab:fdtd} shows achieved LEE and LEE enhancement values for all compositions along with the corresponding NR length and radius values.

\subsubsection{Spectral Overlap Analysis}

\begin{figure}[htbp]
\centering
\begin{subfigure}{0.48\textwidth}
  \centering
  \caption*{(a)}
  \includegraphics[width=\linewidth]{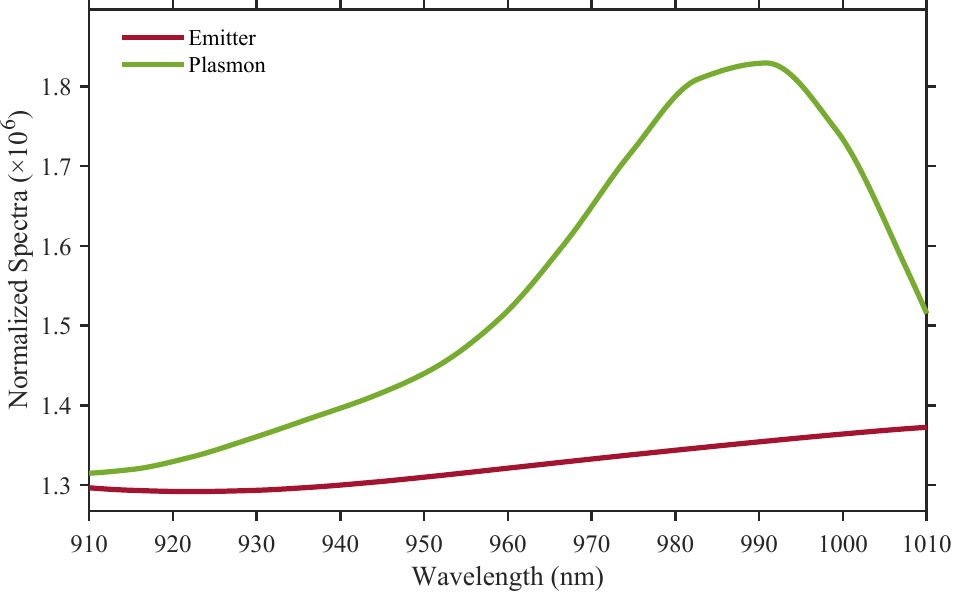}
\end{subfigure}
\hfill
\begin{subfigure}{0.48\textwidth}
  \centering
  \caption*{(b)}
  \includegraphics[width=\linewidth]{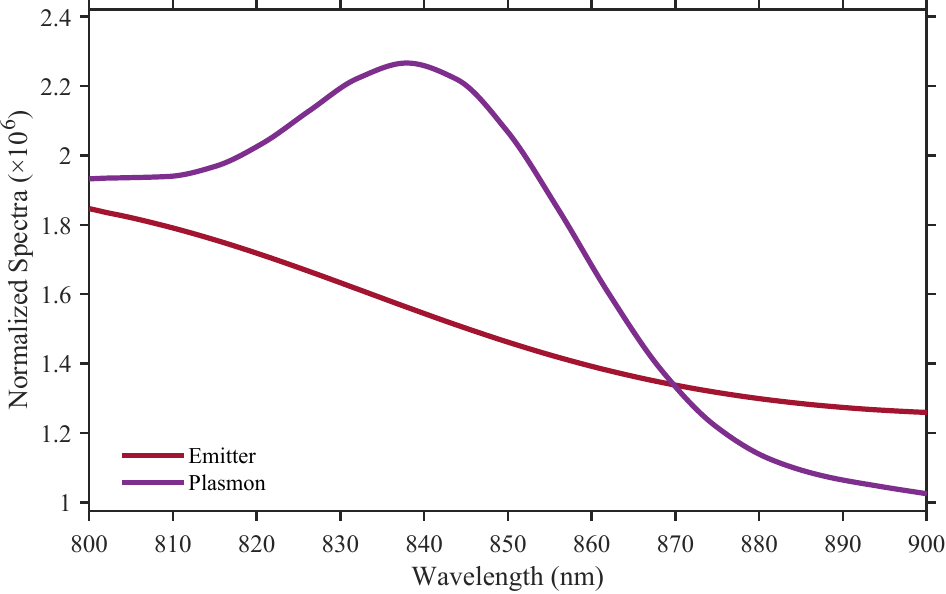}
\end{subfigure}
\begin{subfigure}{0.48\textwidth}
  \centering
  \caption*{(c)}
  \includegraphics[width=\linewidth]{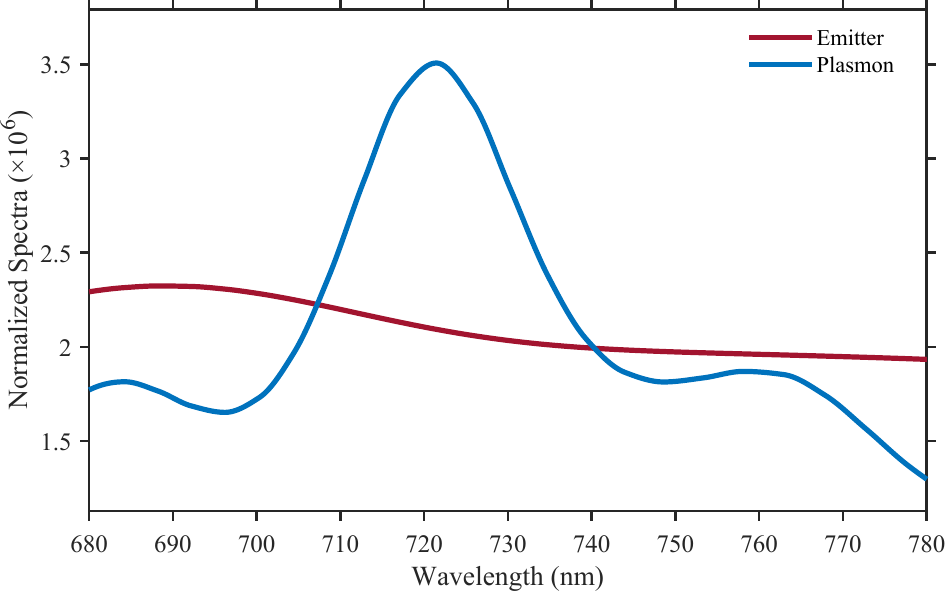}
\end{subfigure}
\hfill
\begin{subfigure}{0.48\textwidth}
  \centering
  \caption*{(d)}
  \includegraphics[width=\linewidth]{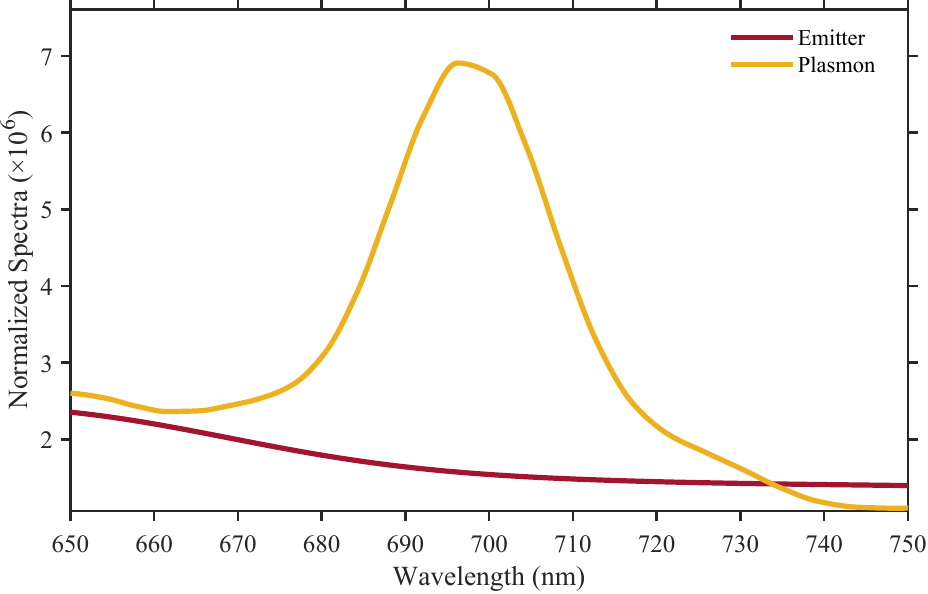}
\end{subfigure}
\begin{subfigure}{0.48\textwidth}
  \centering
  \caption*{(e)}
  \includegraphics[width=\linewidth]{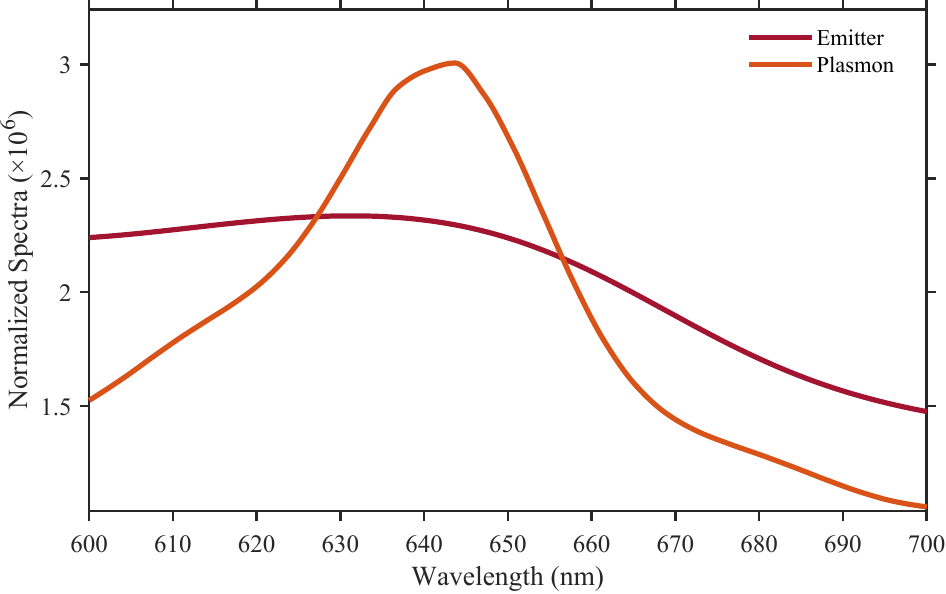}
\end{subfigure}
\caption{Composition-dependent spectral overlap in CsSn$_x$Ge$_{1-x}$I$_3$-based LED for (a) $x = 1$, (b) $x =0.75$, (c) $x =0.5$, (d) $x =0.25$, (e) $x =0$.}
\label{fig:spectral}
\end{figure}

Spectral overlap presents the amount of alignment of the plasmon resonance spectrum with the emitter’s photoluminescence spectrum~\cite{Zhang2014}. Transverse mode LSPR is used which can be effective for spectral overlap alignment in perovskites having refractive index $n \approx 2.0\text{--}2.6$. Composition-dependent LEE and LEE enhancement values for all values of x in CsSn$_x$Ge$_{1-x}$I$_3$ are shown in Fig.~\ref{fig:spectral}. Highest spectral overlap ($J_{\mathrm{cos}}$) of about 96\% has been achieved for both CsSn$_{0.75}$Ge$_{0.25}$I$_3$ and CsSnI$_3$ at the peak Purcell wavelength. $J_{\mathrm{cos}}$ is defined by~\cite{Wurtz2007}.

\begin{equation}
    J = \int S_{\mathrm{emit}}(\lambda)\, C_{\mathrm{plasmon}}(\lambda)\, d\lambda
\end{equation}

\begin{equation}
    J_{\mathrm{cos}} =
    \frac{
        \int S(\lambda)\, C(\lambda)\, d\lambda
    }{
        \sqrt{\int S^{2}(\lambda)\, d\lambda}\;
        \sqrt{\int C^{2}(\lambda)\, d\lambda}
    }
\end{equation}

Achieved $J_{\mathrm{cos}}$ values for all compositions of CsSn$_x$Ge$_{1-x}$I$_3$ and the corresponding NR length and radius values are shown in Table~\ref{tab:fdtd}.

The refractive index of perovskites influences the LSPR spectral position due to changes in the dielectric environment, enabling resonance tuning by design of the nanostructure geometry~\cite{Bayles2020}. This also help in broadening plasmon spectrum that covers a larger portion of the emitter spectrum increasing overlap. Plasmon can boost emission where the perovskite spectrum falls off. This increases overlap even if the peaks are not perfectly aligned~\cite{Kumar2024}. Presence of vertical scattering channels also increases radiative contribution in the transverse mode~\cite{Ooi2024}.

\subsubsection{Far-field Emission Profiles}

Far-field plots depict the amount of light escaping LED and radiating outward as freely propagating radiation~\cite{Mao2021}. Previously discussed mechanisms such as enhancement of near-field to far-field conversion through plasmonic scattering, Purcell-enhanced radiative decay, symmetry breaking of waveguide modes, and transverse LSPR act together to provide strong far-field emission profiles.

\begin{figure}[htbp]
    \centering
    \includegraphics[width=\linewidth]{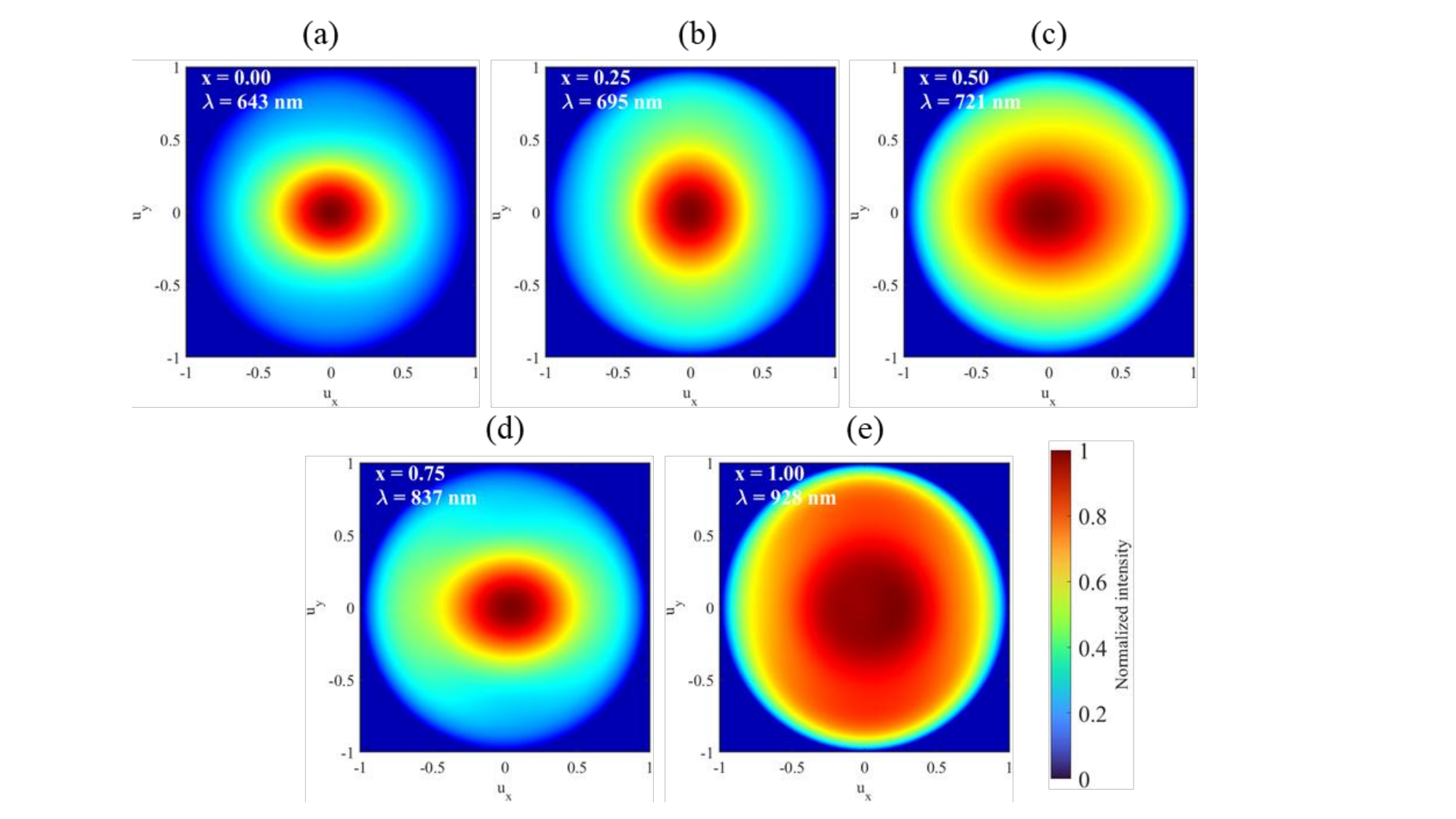}
    \caption{Far-field emission profiles of CsSn$_x$Ge$_{1-x}$I$_3$-based LEDs for (a) $x=1$, (b) $x=0.75$, (c) $x=0.5$, (d) $x=0.25$, and (e) $x=0$. Panel (f) shows the common color scale corresponding to the normalized far-field intensity used for all compositions.}
\label{fig:farfield}
\end{figure}

Composition-dependent far-field plots for all values of x in CsSn$_x$Ge$_{1-x}$I$_3$ are shown in Fig.~\ref{fig:farfield}. Sn-rich ($x = 1$) compositions emit in the deeper NIR. Here the transverse LSPR scatters weakly, giving a dimmer far-field pattern. But as the Ge content increases, the emission shifts to shorter wavelengths. This blue shift is caused by the bandgap widening due to Ge alloying.  These wavelengths couple more strongly to the Au NR resonance, producing stronger scattering and a brighter far-field map. So, the far-field intensity rises steadily with increasing Ge fraction in CsSn$_x$Ge$_{1-x}$I$_3$  based LEDs.

\begin{figure}[htbp]
\centering

\begin{minipage}{0.95\textwidth}
  \centering
  \caption*{(a)}
  \includegraphics[width=\linewidth]{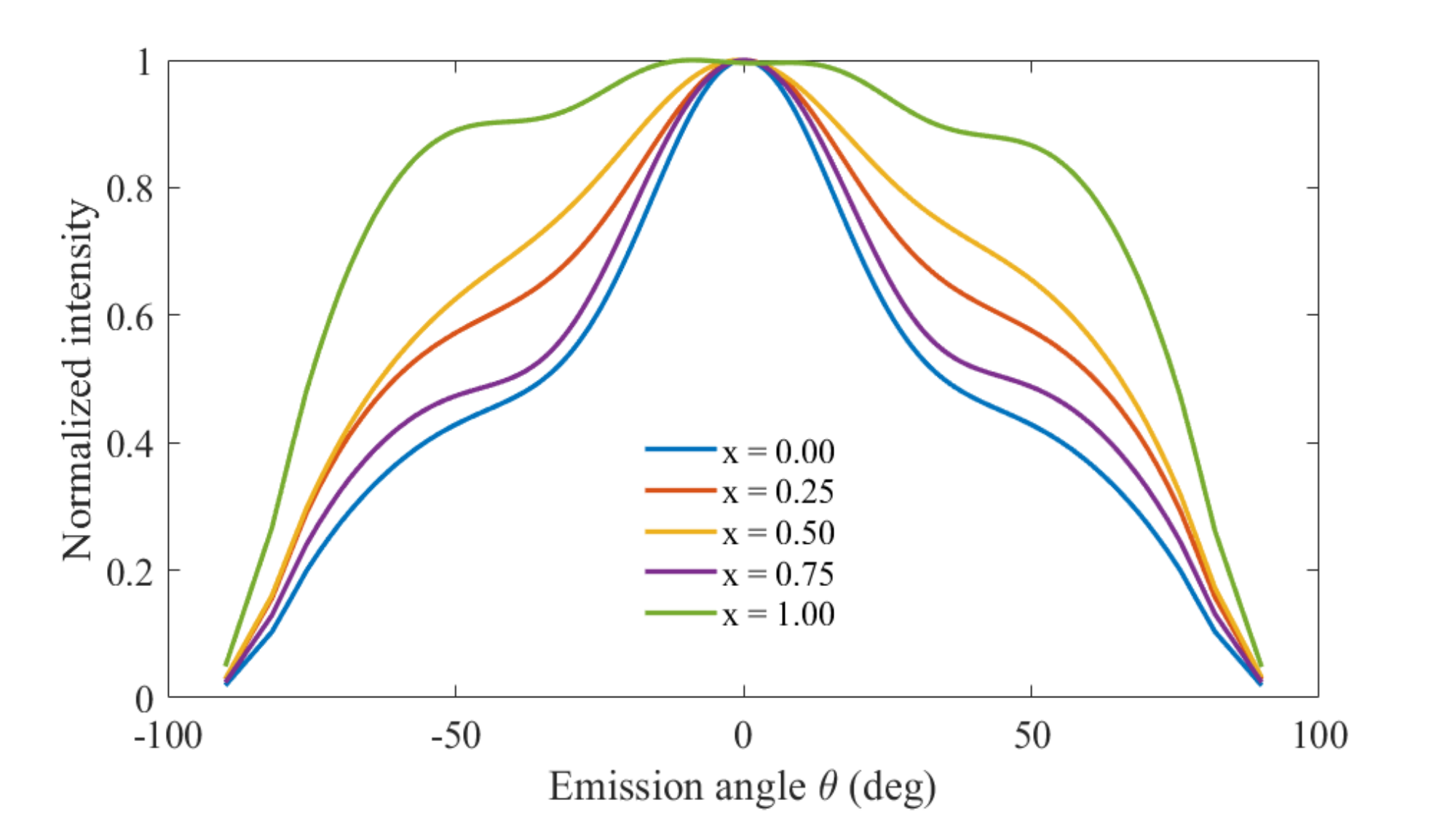}
\end{minipage}

\vspace{4mm}

\begin{minipage}{0.48\textwidth}
  \centering
  \caption*{(b)}
  \includegraphics[width=\linewidth]{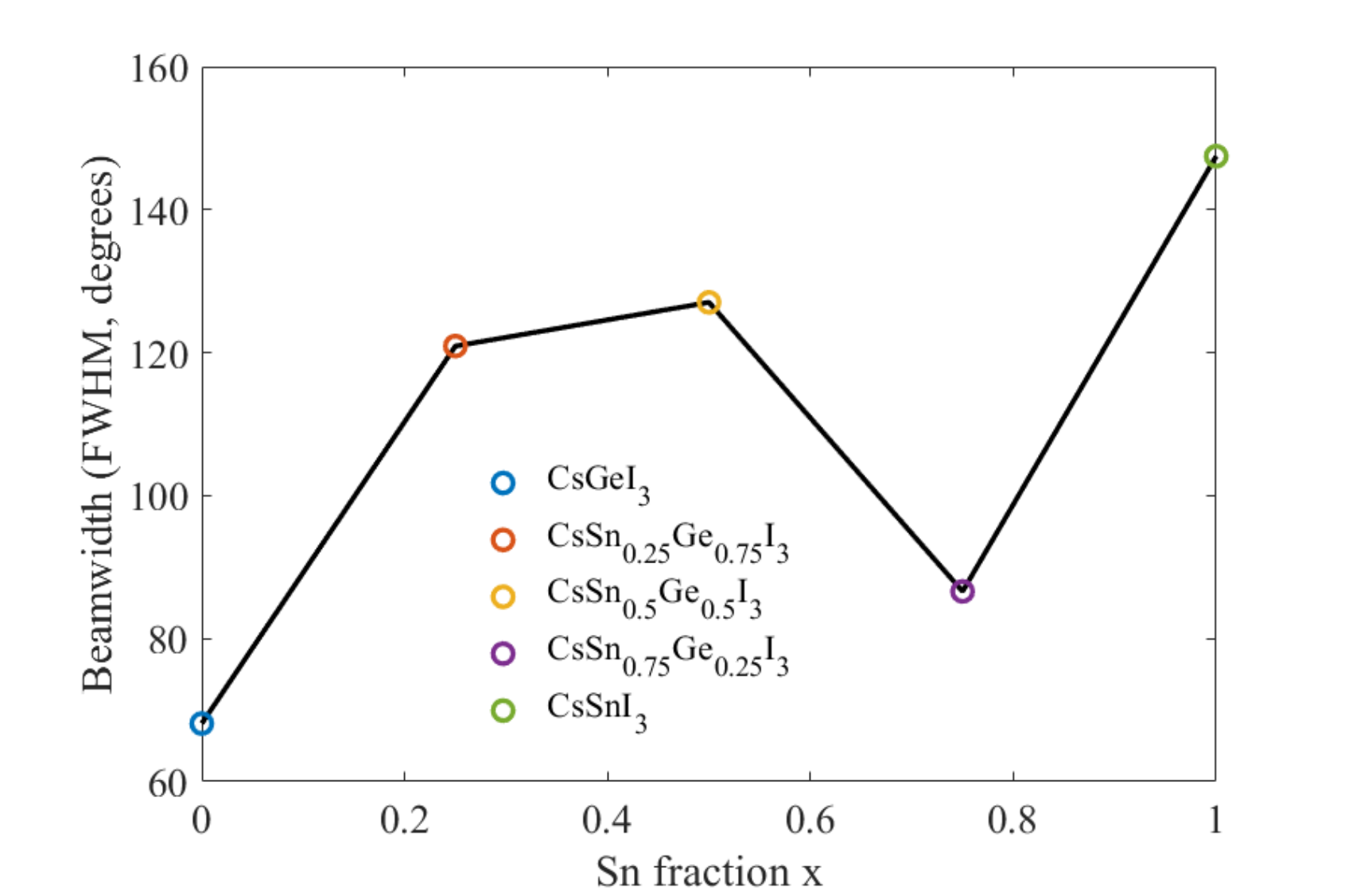}
\end{minipage}
\hfill
\begin{minipage}{0.48\textwidth}
  \centering
  \caption*{(c)}
  \includegraphics[width=\linewidth]{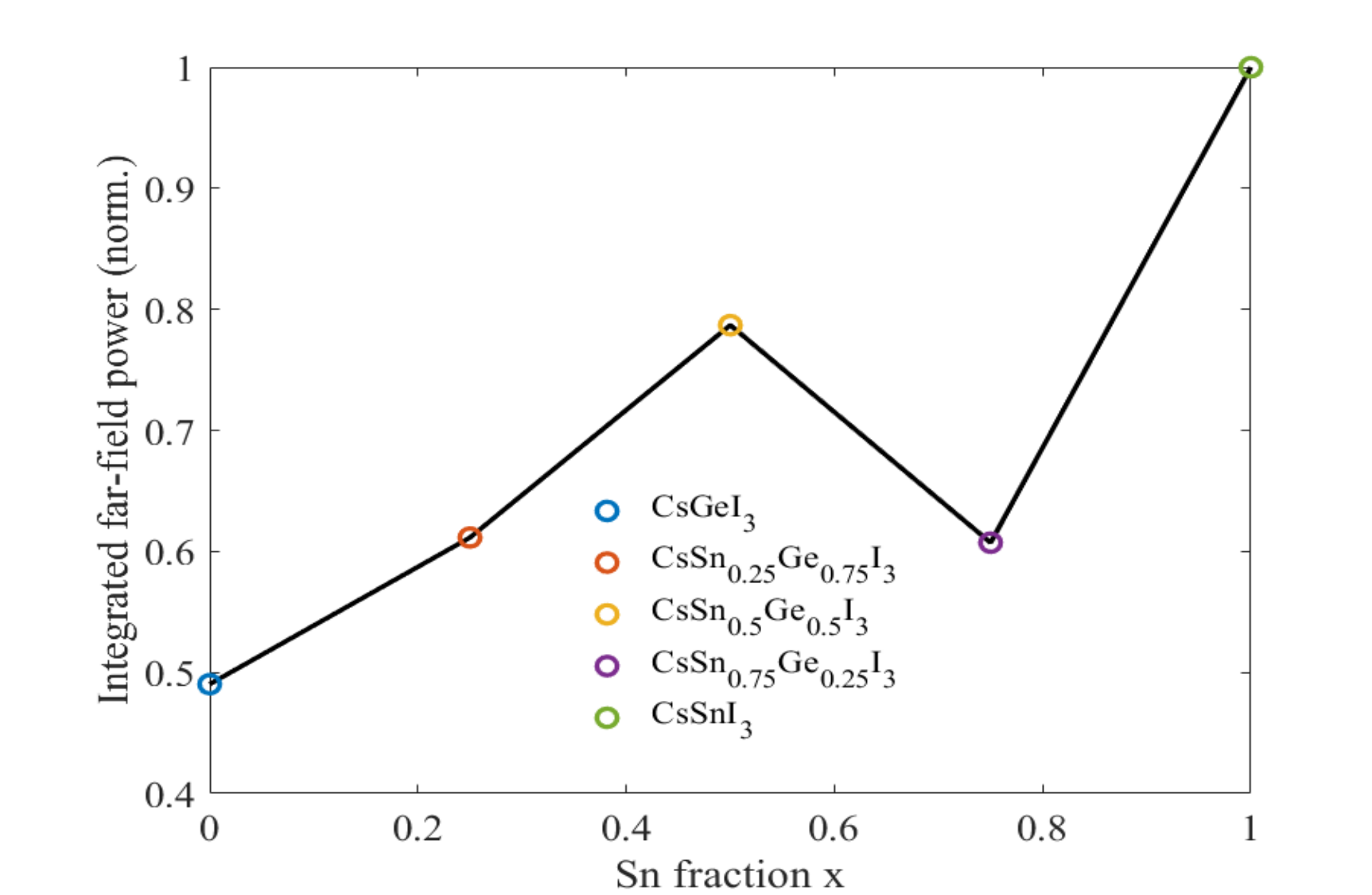}
\end{minipage}

\caption{Far-field emission analysis of CsSn$_x$Ge$_{1-x}$I$_3$-based LEDs. 
(a) Angular line cuts extracted from the far-field emission maps, illustrating composition-dependent emission directionality. 
(b) Beamwidth (FWHM) of the angular emission profiles, quantifying the transition from directional to diffuse emission with increasing Sn content.
(c) Integrated far-field radiated power as a function of Sn composition, highlighting a non-monotonic extraction trend.}

\label{fig:farfield_analysis}
\end{figure}

Figure~\ref{fig:farfield_analysis}a presents the angular line cuts extracted from the far-field emission maps for all CsSn$_x$Ge$_{1-x}$I$_3$ compositions. The emission profiles exhibit strong composition dependence with Ge-rich alloys showing relatively narrow angular distributions. Sn-rich compositions display significantly broadened emission. This broadening is attributed to the higher refractive index of Sn-rich perovskites. This enhancement redistributes radiative power over wider emission angles.

To quantify the total radiative output, the far-field intensity was integrated over all emission angles (Fig.~\ref{fig:farfield_analysis}b). The integrated power exhibits a non-monotonic dependence on composition, reaching a maximum near $x = 0.5$. This indicates that optimal far-field extraction does not coincide with the strongest Purcell enhancement but instead arises from a balance between emission rate enhancement and angular outcoupling.

Figure~\ref{fig:farfield_analysis}c summarizes the angular emission characteristics using the FWHM of the angular profiles. The FWHM increases with Sn content. This confirms the transition from directional emission in Ge-rich compositions to more diffused radiation in Sn-rich alloys.

\subsection{Comparison with reported PeLED architectures}

To place the performance of the CsSn$_x$Ge$_{1-x}$I$_3$-based PeLEDs in context, Table~\ref{tab:comparison} compares the Purcell factor and LEE reported for perovskite LED architectures published between 2021 and 2025. The table focuses mainly on lead-free and reduced-lead systems. Lead-based CsPbBr$_3$ devices are included only as reference benchmarks. Many earlier studies improve either the emission rate or the optical outcoupling, but not both at the same time. Clear and combined reporting of Purcell enhancement and LEE is still uncommon,
especially for lead-free emitters.

\begin{table}[htbp]
\centering
\caption{Comparison of Purcell factor and light extraction efficiency (LEE) in
perovskite LEDs reported between 2021 and 2025. Lead-based CsPbBr$_3$ devices
are included only as benchmarks.}
\label{tab:comparison}
\footnotesize
\setlength{\tabcolsep}{2.5pt}
\renewcommand{\arraystretch}{1.15}

\begin{tabularx}{\linewidth}{Y c c c Y Y c}
\hline
Emitting material & Emission (nm) & Purcell factor & LEE / outcoupling &
Optical structure & Methodology & Ref. \\
\hline

CsSn$_{0.25}$Ge$_{0.75}$I$_3$ (this work) &
$\sim$670 &
12$\times$ (sim.) &
$\sim$19\% &
Au/SiO$_2$ nanorod &
DFT $n,k$ + FDTD &
This work \\

CsSn$_{0.5}$Ge$_{0.5}$I$_3$ (this work) &
$\sim$730 &
5.3$\times$ (sim.) &
25\% &
Au/SiO$_2$ nanorod &
FDTD, near--far field &
This work \\

FASnI$_3$ &
$\sim$880 &
PF inferred &
Grating-assisted &
DFB grating &
Exp + mode analysis &
\cite{MartinezPastor2025} \\

Cs$_2$AgBiBr$_6$ &
NIR &
-- &
$\sim$42\% (sim.) &
Metal microcavity &
FDTD &
\cite{Tabibifar2024} \\

Cs$_3$Cu$_2$I$_5$ &
$\sim$440 &
Implicit &
Directional &
VCSEL cavity &
Experiment &
\cite{Li2023Cu} \\

CsPbBr$_3$ (bench.) &
$\sim$530 &
$\sim$2--3$\times$ &
31--38\% &
Planar microcavity &
TMM + dipole &
\cite{Lin2021Nano} \\

CsPbBr$_3$ (bench.) &
$\sim$520 &
$\sim$2$\times$ &
$\sim$18$\times$ &
Mie resonator &
FDTD + TRPL &
\cite{He2023Nanoscale} \\

\hline
\end{tabularx}
\end{table}

As shown in Table~\ref{tab:comparison}, this work is among the few studies that quantitatively evaluate both Purcell enhancement and LEE within a single lead-free
device platform. When compared with CsPbBr$_3$ microcavity and resonator benchmarks, the CsSn$_{0.5}$Ge$_{0.5}$I$_3$ composition shows competitive light extraction. At the same time, it maintains strong emission-rate enhancement. These results highlight the benefit of using composition-aware plasmonic design to balance light generation and light extraction.

\subsection{Design Considerations}

The best Purcell factor does not necessarily lead to the best light extraction. This is due to a clear tradeoff between near-field enhancement and outcoupling. At $x = 0.25$, the emitter spectrum and the plasmon resonance are almost perfectly aligned. This strong match accelerates photon generation and drives the Purcell factor up to about 12$\times$. The drawback is that this composition has a relatively high refractive index of roughly 2.5, which traps much of the emitted light inside the device.

\begin{figure}[htbp]
    \centering
    \includegraphics[width=\linewidth]{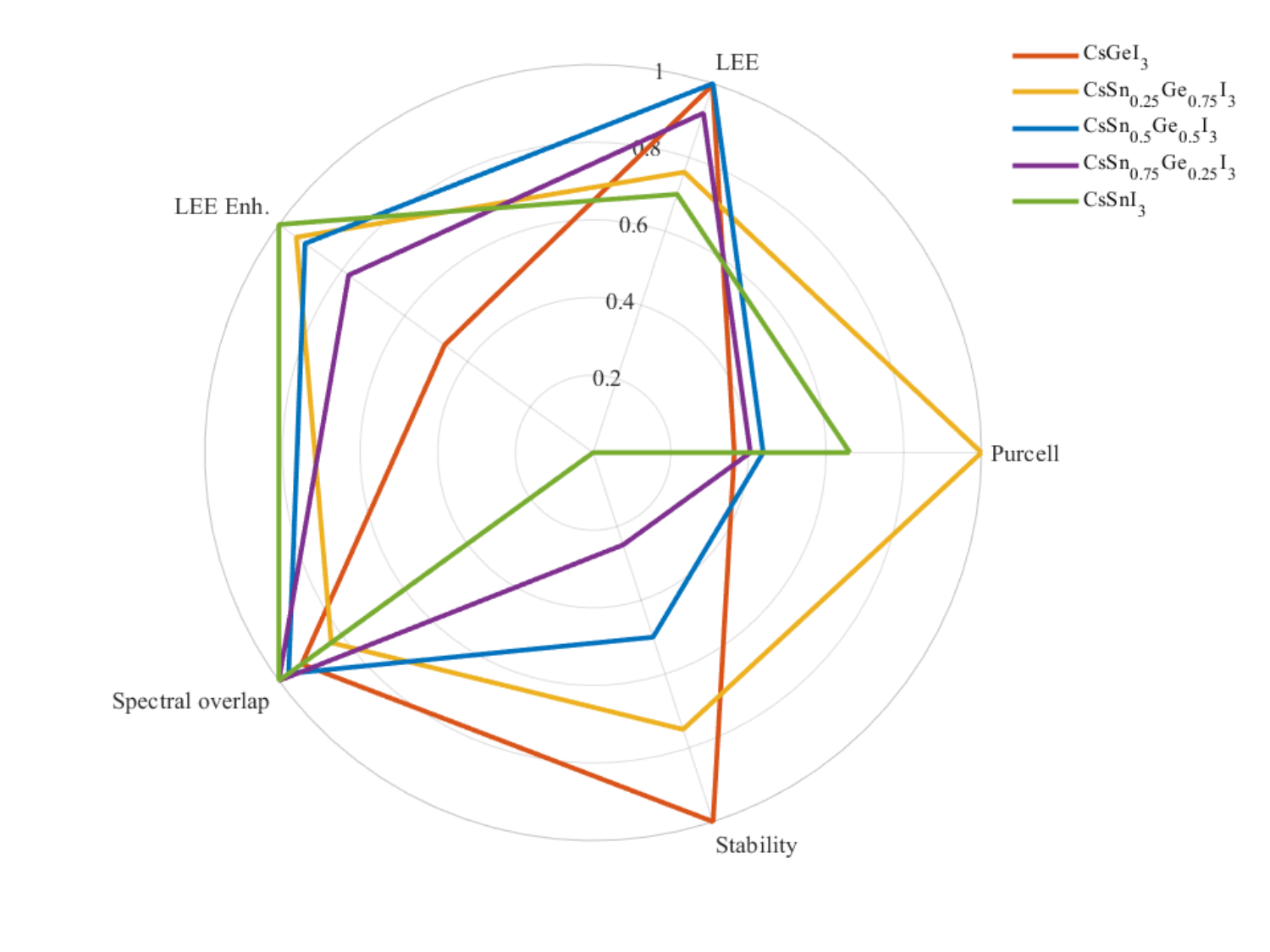}
    \caption{Spider plot comparing normalized performance metrics of CsSn$_x$Ge$_{1-x}$I$_3$ PeLEDs, including Purcell factor, LEE, LEE enhancement, spectral overlap, and compositional stability. Compositional stability is represented using the Ge fraction $(1-x)$ as a qualitative proxy.}
    \label{fig:design_spider}
\end{figure}
\begin{figure}[htbp]
\centering
\begin{subfigure}{0.75\textwidth}
  \centering
  \caption*{(a)}
  \includegraphics[width=\linewidth]{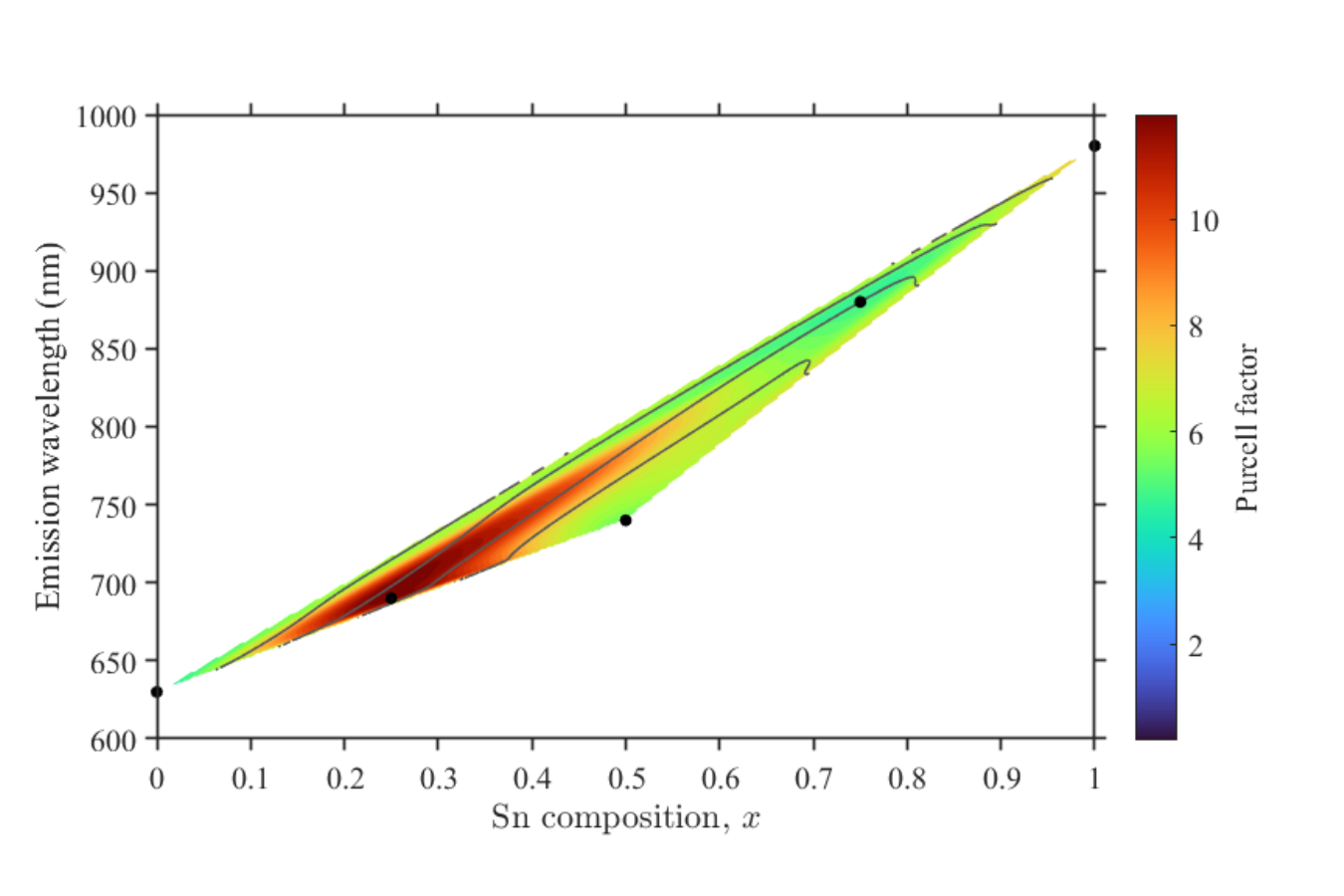}
\end{subfigure}
\hfill
\begin{subfigure}{0.75\textwidth}
  \centering
  \caption*{(b)}
  \includegraphics[width=\linewidth]{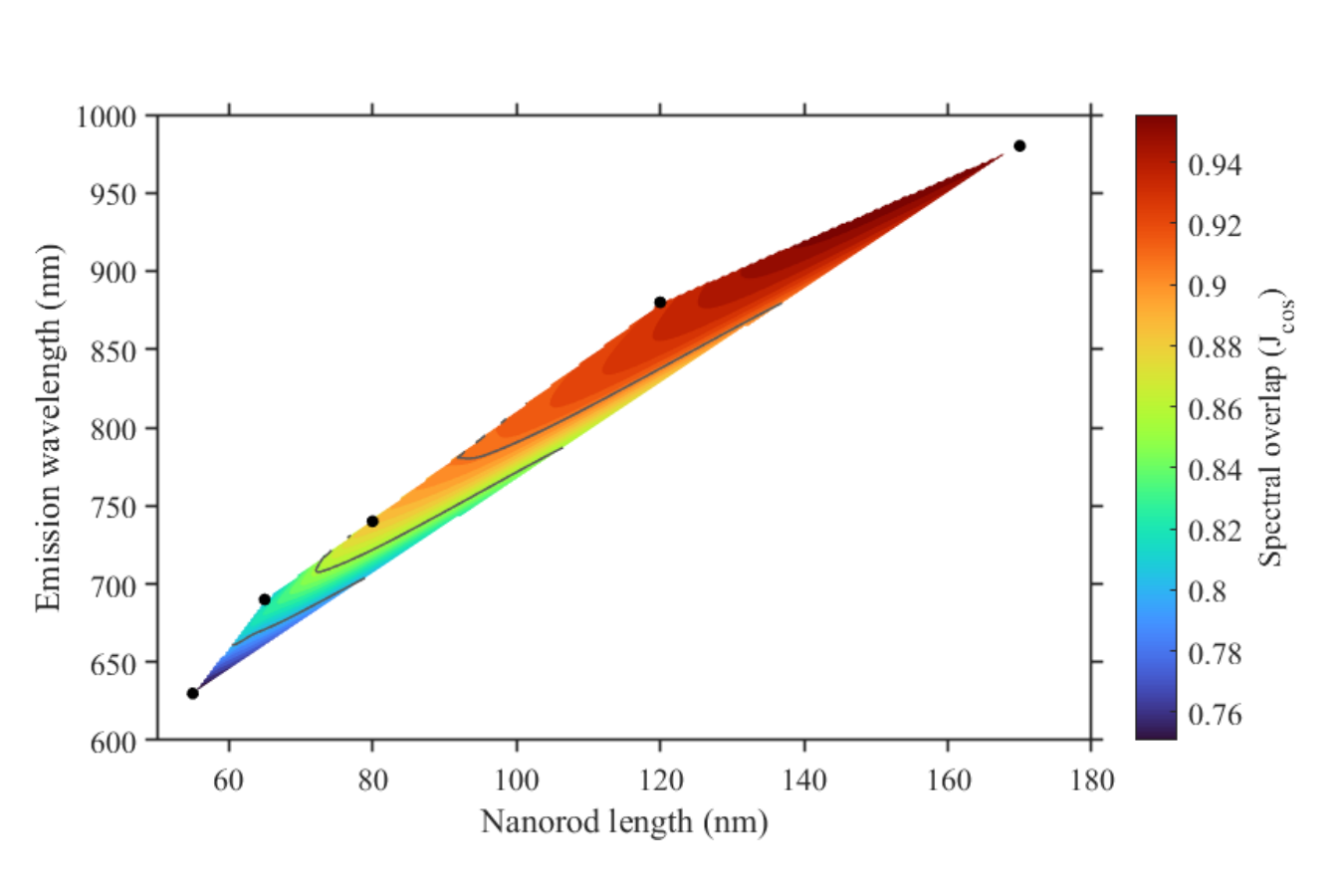}
\end{subfigure}

    \caption{Design landscapes illustrating competing emission and extraction mechanisms in CsSn$_x$Ge$_{1-x}$I$_3$ PeLEDs. 
    (a) Purcell--LEE tradeoff map as a function of Sn composition $x$ and emission wavelength, where the color scale denotes the Purcell factor and contour lines indicate regions of comparable light-extraction efficiency. 
    (b) Spectral overlap landscape ($J_{\mathrm{cos}}$) between the emitter spectrum and plasmonic resonance as a function of nanorod length and emission wavelength, showing a systematic shift of optimal overlap toward longer nanorods for Sn-rich compositions.}
    \label{fig:design_maps}
\end{figure}

At $x = 0.5$, the spectral match is slightly weaker, but the lower refractive index reduces total internal reflection, allowing more photons to escape. In practice, this means that $x = 0.25$ is more efficient at producing light, while $x = 0.5$ is more efficient at releasing it. This distinction shows that Purcell enhancement and far-field extraction should be treated as two separate but equally important design factors. Simply increasing the spontaneous emission rate does not automatically lead to the best overall device performance.

These trends also guide the choice of nanorod geometry. Sn-rich compositions emit farther into the NIR and therefore couple best to longer nanorods of about 170 nm, which match the red-shifted plasmon resonance. Ge-rich compositions emit at shorter wavelengths and achieve stronger coupling with shorter nanorods in the range of 55 to 70 nm.

These competing effects are summarized in Fig.~\ref{fig:design_spider}, which compares Purcell enhancement, spectral overlap, light extraction efficiency, and LEE enhancement across compositions. To provide a qualitative indication of material robustness, compositional stability is represented using the Ge fraction $(1-x)$ as a proxy. This reflects the improved resistance to Sn-related degradation in Ge-rich alloys.

While Fig.~\ref{fig:design_spider} highlights discrete composition-dependent trends, the underlying tradeoffs can be more clearly visualized using continuous design maps. Figure~\ref{fig:design_maps} presents two-dimensional landscapes that show how Purcell enhancement, light extraction, and spectral overlap evolve simultaneously with composition, emission wavelength, and plasmonic geometry. These maps complement the spider plot by revealing optimal design regions rather than isolated operating points.

\section{Conclusion}

In this work we used atomistic simulation tools to calculate optical and electrical property of material and then utilized FDTD simulations to establish a practical design framework for lead-free CsSn$_x$Ge$_{1-x}$I$_3$ perovskite LEDs with plasmonic enhancement. Among all tested compositions, CsSn$_{0.5}$Ge$_{0.5}$I$_3$ delivers the most balanced performance. It achieves the highest light extraction efficiency (25\%), strong Purcell enhancement (5.3$\times$), 93\% spectral overlap, and improved oxidation resistance from the native GeO$_x$ passivation layer \cite{Chen2018}. For applications prioritizing spontaneous emission rate, CsSn$_{0.25}$Ge$_{0.75}$I$_3$ offers the largest Purcell enhancement (12$\times$). The results also suggest that longer nanorods suit Sn-rich NIR emitters while shorter nanorods suit Ge-rich visible emitters.

The simulations do not yet include lossy optical modes or non-radiative recombination, so actual device efficiencies may differ. Experimental validation with fabricated CsSn$_x$Ge$_{1-x}$I$_3$ LEDs and embedded plasmonic nanostructures is the natural next step.

Overall, the combination of tunable emission, strong plasmonic response, and Ge-induced stability makes CsSn$_x$Ge$_{1-x}$I$_3$ a compelling lead-free candidate for flexible and wearable optoelectronics.

\begin{backmatter}

\bmsection{Funding}

\bmsection{Acknowledgment}
The authors acknowledge the support received from the Department of Electrical and Electronic Engineering, Bangladesh University of Engineering and Technology.

\bmsection{Disclosures}
The authors declare no conflicts of interest.

\bmsection{Data Availability Statement}
Data underlying the results presented in this paper are not publicly available at this time but may be obtained from the authors upon reasonable request.

\end{backmatter}

\bibliography{references}

\end{document}